\documentclass{aa} 

\usepackage{graphicx}
\usepackage{txfonts}
\usepackage{subfig}
\usepackage{threeparttable}
\usepackage{amsmath}
\usepackage{mathtools}

\newif\ifhidealgebra
\hidealgebratrue 
\newcommand{\algebra}[1]{\ifhidealgebra\else\textcolor{gray}{\begin{align*}{#1}\end{align*}}\fi}

\begin{document}

    \title{A broken debris cascade as a possible source of hot dust emission in transitioning planet-forming disks}
    \titlerunning{A broken debris cascade in transitioning disks}

   \author{N. Swinkels 
   \and C. Dominik}
   
   \institute{Anton Pannekoek Institute for Astronomy, University of Amsterdam, Science Park 904, 1098 XH Amsterdam, The Netherlands \\
              \email{niels.swinkels2@student.uva.nl}
              and
              \email{dominik@uva.nl}
             }
   \date{Received March 12, 2024; accepted April 24, 2024}

  \abstract
   {Planet-forming disks go from gas-rich, massive disks made of dust and gas into planetary systems containing only small amounts dust. This dust is produced by collisions between smaller planetary objects, such as planetesimals, asteroids, and comets. Traditionally, we talk about protoplanetary (age $\sim 1$\,Myr), transitional ($\sim$ 5--10\,Myr), and debris disks ($\sim$ 10-hundreds of\,Myr) even though the overlap between these phases may be relevant.}
   {We aim to show that in the transition phase of a disk, when the gas surface densities are reduced but not yet negligible, a seemingly small amount of collisional activity may lead to the production of dust on a level that is observationally relevant by creating regions characterised by an optical depth of 1 or above. In particular, we aim to show that the hot dust emission component of transitional disks may in fact be debris dust that has been produced in such collisions.}
   {We developed an analytical model to derive the conditions in which observationally relevant amounts of dust can be produced. We focussed on the effect of the gas surface density during the transition phase of a disk from fully gas-dominated to the gas-poor debris stage.}
   {We show that the decrease in the gas surface density has an important effect.  It allows for smaller planetesimals to become collisional, initiating a cascade. At the same time, small particles are not destroyed by collisions. This interrupted cascade is critical in terms of preserving the produced dust for significant time intervals, allowing for a seemingly minor amount of planetesimal collisions to be effective in producing detectable amounts of dust.}
   {The warm emission of low amounts of dust in many transitional planet-forming disks might be caused entirely by second-generation dust, exposing planetary material at a much earlier time  than the age that debris disks have traditionally
been characterised by.}

   \keywords{Protoplanetary disks, Planets:formation, Infrared: stars}

   \maketitle
%

\section{Introduction}
The classical picture of the formation of a planetary systems divides the process into a number of steps: (i) the collapse of a molecular core in a molecular cloud, (ii) disk formation due to conservation of angular momentum during collapse as well as due to viscous spreading, (iii) planet formation in the disk by core-accretion and possibly gravitational instabilities, (iv) clearing out the gas in the disk by a combination of viscous accretion onto the star, planet formation, and photoevaporative and/or magnetically driven outflows (winds), and (v) mature planetary system in which rings of planetesimals can produce dust in collisions, ushering in the debris disk phase of the disk, which can last for gigayear scales.

Over the years, it has become clear that these processes are not nearly as sequential and ordered as this simple picture suggests. Very young disks  already contain structures that have been interpreted in the context of embedded planets \citep{ALMA-2015-HLTau,Valegard-2021-IMTT}. Theoretical models of planet formation show the process starting early and continuing over extended periods of time \citep{Drazkowska-PPVII}. Debris disks can be fully developed at quite young ages of only 10 or 20 Myrs \citep{Lagrange-PPVI}.

One of the more puzzling observations in young planet-forming disks is the observation of a significant component of near infrared (NIR) emission that is attributed to very warm dust near typical sublimation temperatures around 1500\,K. This emission is indeed located close to the star \citep[e.g.,][]{Benisty_2011}. It was originally interpreted as being due to a disk inner region heated by accretion 
\citep{Hillenbrand-1992,dAlessio-1998-irradiated-disks,dAlessio-accretion-for-excess-1999}. However,  \citet{Natta-2001-inner-disk-edge} and \citet{DDN-2003} realised, that the inner edge of a disk is much more intensely irradiated by that star than a disk surface, leading to enhanced emission also in disks that are heated solely by stellar irradiation.  Even though a full recovery of the measured NIR emission requires enhanced scale heights in the inner disk, the basic assumption has been that these processes are responsible for the observed emission.

\citet{Meeus-2001-groups} established that there are two basic types
of spectral energy distributions (SED) for young Herbig disks that are
distinguished by the amount of mid-IR emission, but that also show
quite similar properties in the NIR.  The assumption stated that these
disks differ in the amount of flaring (group-I versus group II disks,
\citet{DD-2004-flaring-shadow}), but that the properties of the
innermost edge of the disk at the sublimation point are very similar,
with the IR emission amounting to 10-30\% of the stellar flux
\citep{Dominik-2003-passive-disk-model}. However, eventually it became
clear that the group-I disk SED is a symptom of a large inner hole in
the disk \citep{Maaskant-2013-group-I-transitional}; this hole is akin
to the one we would expect to be carved out by a forming planetary
system. The basic picture is then that a massive planet creates a dust trap in a pressure bump outside the planetary orbit, keeping most of the dust from flowing into the inner disk region.  Gas can still pass the planetary orbit through streamers, but the flow of dust would be substantially reduced. This then throws some doubt on the interpretation of the NIR excess.  Consequently, the question arises of why, even though there is a nearly empty cavity,  the dust suddenly re-appears at the inner edge, close to the star.  In viscous accretion disks, the accretion timescales are shortest in the inner disk and we would expect the hole to be largely emptied out.

A number of hypotheses have been put forward to explain the detected emission, even in the presence of a strongly dust-depleted gap. \citet{Casassus-2013-streamers} detected fast streamers in the cavity of HD142527.  These streamers can be created through the interaction with the observed companion of that star \citep{price-2018-streamers}.  If these streamers are narrow and fast, this may account for the low emission coming from the region between inner and outer disk. \citet{Pinilla-2016-tunnel} put forward the idea that small dust grains may sneak through the trap, quickly grow again to larger sizes that are harder to detect, and then fall apart at the snowline.  There the water ice that may have been binding dust aggregates evaporates, leading to the sudden appearance of small grains that drift more slowly. The resulting traffic jam is then responsible for the observed NIR emission.

In this paper, we explore the idea that the dust seen at the inner edge of these transition disks is not primary dust funneled from the dust trap into the inner system, but that it is secondary dust, produced by the collisions of planetesimals.  This would give these disks (at least partially) the status of debris disks.

There is a large body of literature on dust production in debris disks. In addition to a number of analytical estimates and computations \citep[e.g.,][]{Wyatt_Dent_2002,vega_phenomenon,2006A&A...455..509K,2007ApJ...663..365W,2007ApJ...658..569W,2008ApJ...673.1123L,2010Icar..206..735K,2010MNRAS.405.1253K,2011CeMDA.111....1W}, numerical modeling of debris disks over long times is an established field \citep[e.g.,][]{2003A&A...408..775T,Krivov_basics_2005,Krivov_model_2006,2007A&A...472..169T,2014A&A...571A..51V}. Other works have put the production of debris into the context of a model starting with a gas-rich disk, focussing on the effect of the gas on the larger parent bodies for dust production \citep[e.g.,][]{2005AJ....130..269K,2008ApJS..179..451K,2010ApJS..188..242K,2011A&A...530A..62R,2012A&A...541A..11R,2014MNRAS.442.3266K}.
The onset of debris formation in gas-rich disks during planet formation has been proposed in a set of two papers focussed on the disk of HD\,163296 \citep{Turrini-2019T,dAngelo-Marzari-2022} and studied further in a parameter study \citep{Bernabo_2022}. These are very detailed models that consider the scattering of planetesimals by the forming giant planets in the system, and they estimate the amount of dust production resulting from these collisions, zooming in on a specific disk.

In the present paper, we are asking a more focussed and basic question, namely: what conditions are needed in order to create a torus of moderate optical depth in the inner regions of a disk required to reproduce a low-mass, high scale-height inner disk.  We achieved this by developing an analytical model that captures the dust production of a cascade in the presence of gas, while considering that the gas may stop the collisional cascade for the smallest particles. In particular, we discuss the relevance of a reduced gas surface density to constrain when such a broken cascade can occur and consider the relation between the mass input from collisions and the optical depth of the torus.

In Sect.~\ref{sec:gas_effect}, we show the effect of drag on the collisional cascade, derive the important timescales associated with it, and show how it breaks the cascade. These effects are applied in Sect.~\ref{sec:broken_cascade} to derive the size distribution of particles in the system. In Sect.~\ref{sec:required_mass_flux}, we calculate the amount of particles needed to create a moderately optically thick cloud of dust; alongside it, we show how much mass flux of incoming planetesimals is needed to sustain such a structure. In Sect.~\ref{sec:results}, we evaluate the derived equations, derive and discuss the outcome and put it into context.  Finally, in Sect.~\ref{sec:conclusions}, we summarise our conclusions.

\section{Effect of gas on the cascade}
\label{sec:gas_effect}
\subsection{Introduction}

\begin{figure}
    \centering
    \includegraphics[width=\linewidth]{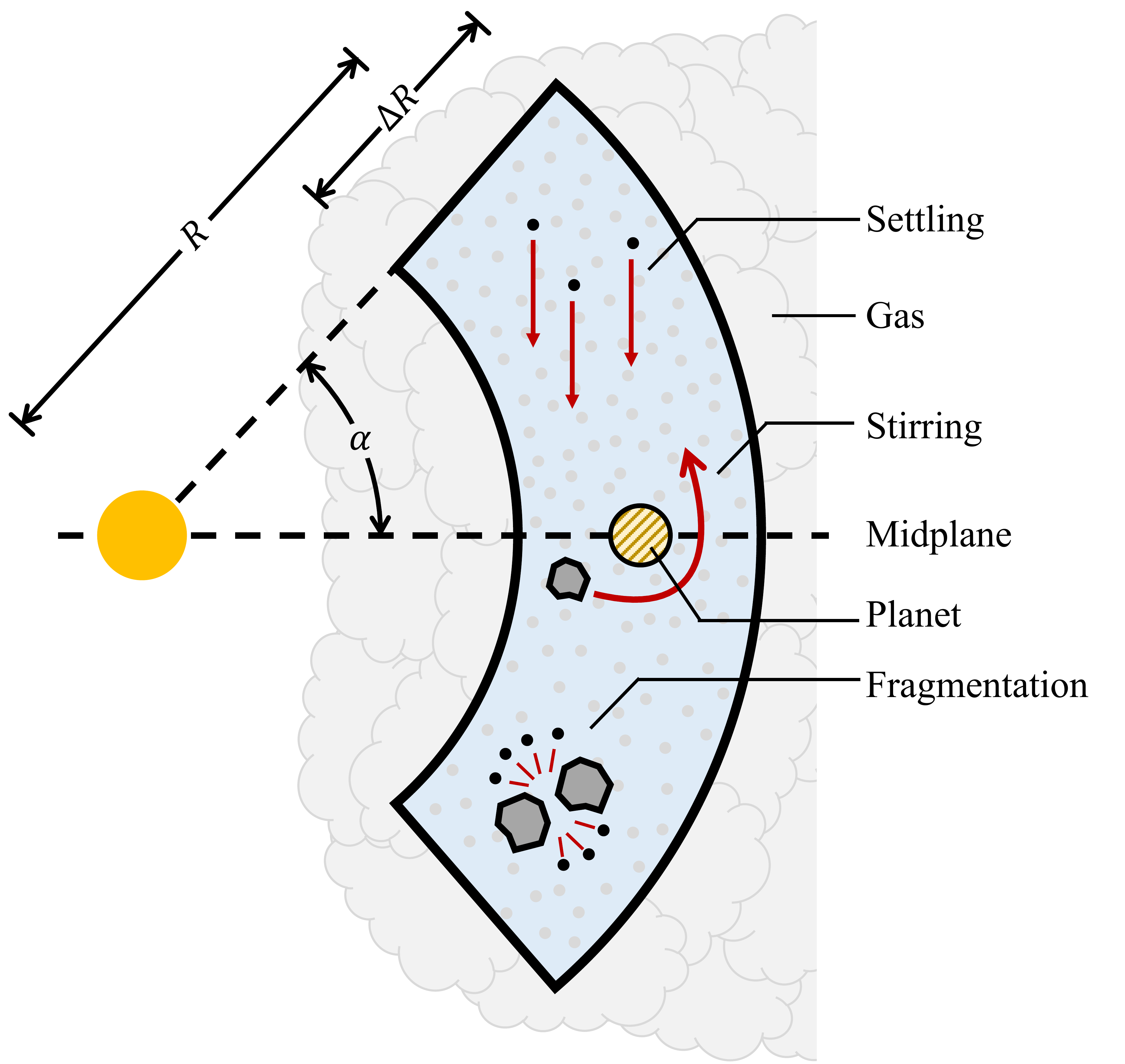}
    \caption{Proposed model consists of a dust cloud located a distance, $R,$ from the star (with a width $\Delta R$), suspended in a gaseous medium. A planet scatters planetesimals out of the midplane, which collide and fragment. The smallest particles are bound to the gas and slowly settle into the midplane.}
    \label{fig:system}
\end{figure}
Figure \ref{fig:system} shows the basic model: close in to the star, a cloud of dust suspended in the disk gas covers a large solid angle $\alpha$ around the star. We assume that we are dealing with a torus-like structure that contains a mix of gas, dust, and planetesimals. A planet scatters planetesimals into highly inclined orbits, which start to collide and fragment into smaller and smaller particles. Particles below a certain size, $s_\mathrm{b}$, are tightly bound (coupled) to the gas and start to move along with it, removing relative velocities between them, preventing further fragmentation. Instead, these particles slowly settle into the midplane. Once located in the midplane, we assume that these particles no longer contribute to the system.  Midplane particles are basically irrelevant for the amount of stellar radiation reprocessing, because the maximum amount of stellar radiation they could absorb even under optimal conditions (optically thick geometrically thin midplane of dust in an otherwise optically thin disk) becomes extremely small unless the disk extends all the way to the star. For an inner disk radius of 0.1 au, that fraction is about 2\%. For an inner radius of 1\,au (close to the sublimation radius in the disks of Herbig stars), only $10^{-3}$ of the stellar photons are absorbed in the midplane \citep{Natta_2001}. These numbers are too small compared to the tens of percent we are looking for. We also assume that these particles become irrelevant for the collisional cascade as they may re-coagulate and drift into the sublimation zone.

\begin{figure}
    \centering
    \includegraphics[width=0.9\linewidth]{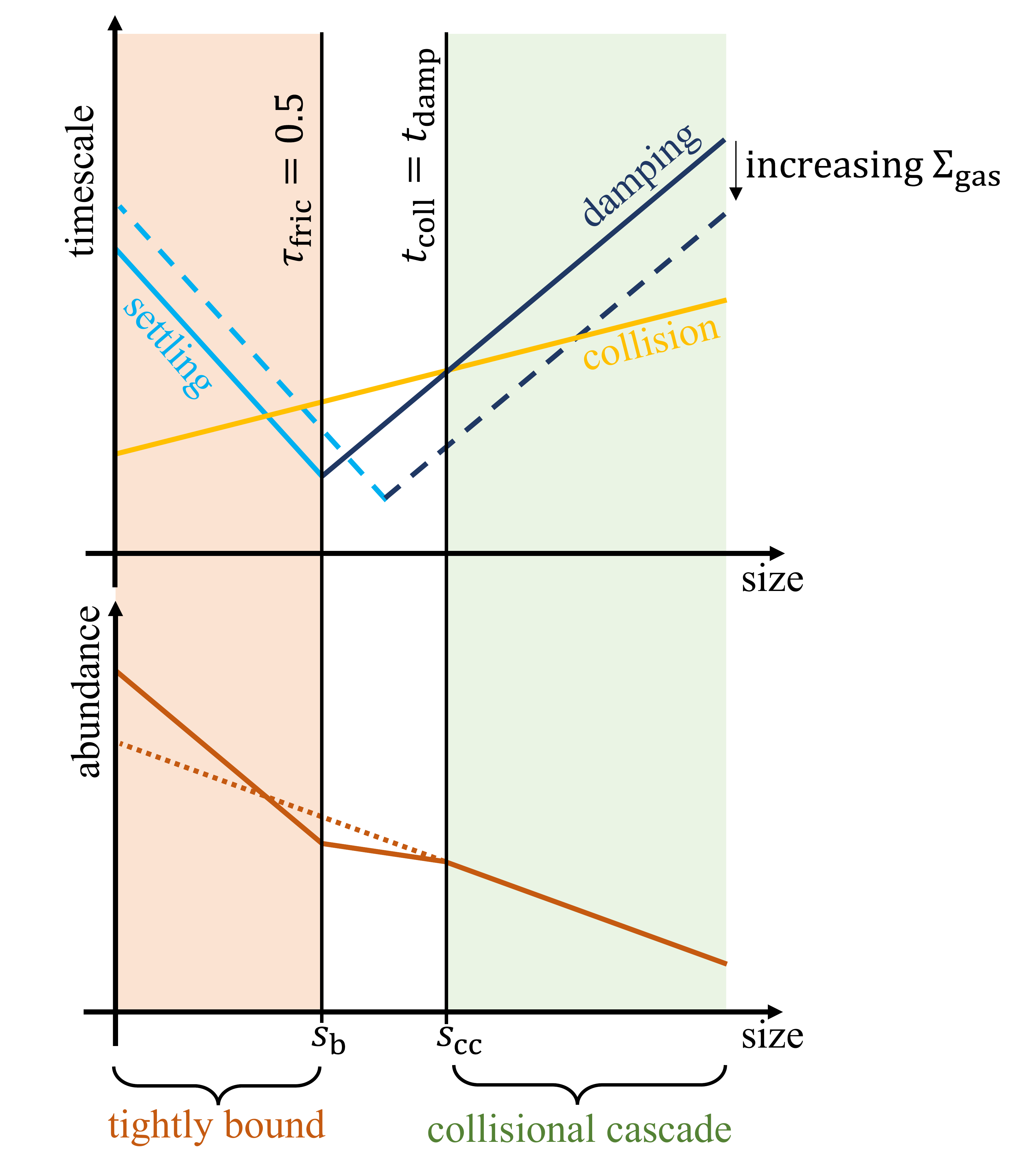}
    \caption{Schematic log-log representation of the timescales governing particles of a certain size (top of the plot) and of the resulting abundances (bottom) of particles at a given size. These abundances are integrated over the torus, but excluding particles that have fully settled to the midplane and are treated as irrelevant. The collisional lifetime (yellow line) of a particle increases with size. Particles smaller than $s_\mathrm{b}$ are tightly bound to the gas, and are dragged into the midplane by gravity on a settling timescale (light blue) which decreases with particle size. For larger particles, the inclination of their orbit is damped on a damping timescale (dark blue), which increases with particle size. Particles smaller than $s_\mathrm{cc}$ have a damping timescale shorter than their collisional timescale, removing them from the collisional cascade. The effect of increasing gas density on these timescales is shown by the dashed lines.}
    \label{fig:timescales_sketch}
\end{figure}

While the smallest particles below $s_\mathrm{b}$ couple to the gas and completely leave their inclined orbits around the star, larger particles retain their orbit. The effect of gas does however slowly dampen their inclination, moving them into the midplane as well. There is some range in the particle sizes which do not couple to the gas, but for which damping occurs quicker than collisions, essentially removing them from the cascade as well. The size above which the collisional timescale is lower, and particles are in the collisional cascade we refer to as $s_\mathrm{cc}$. These timescales and boundary sizes are shown in Fig. \ref{fig:timescales_sketch} and their derivation is described below.

Since the largest particles are constantly fragmenting, replenishment is needed to sustain such a structure over a longer period. This replenishment is provided by the young planet scattering planetesimals into highly inclined orbits. When settling is a much slower process than collisions, the presence of gas will cause an accumulation of small particles, reducing the required mass flux of planetesimals.

\subsection{Coupling, settling, and damping}
To derive what particles are bound to the gas, we consider the Epstein drag, given by \citet{armitage} as:
\begin{equation}
\label{eq:epstein}
    F_\mathrm{D}=\frac{4\pi}{3}\rho_\mathrm{gas}s^2v_\mathrm{th}v,
\end{equation}
with $\rho_\mathrm{gas}$ the gas density, $s$ the particle size, $v$ the particle velocity relative to the gas, and the thermal velocity of gas molecules \citep{armitage} given as:\ \begin{equation}
\label{eq:v_th}
    v_\mathrm{th}=\sqrt{\frac{8k_\mathrm{B}T_\mathrm{gas}}{\pi \mu_\mathrm{mol}m_\mathrm{p}}}.
\end{equation}
The timescale over which the gas significantly affects a particle's motion, the friction-time, is given by \citet{armitage} as:
\begin{equation}
\label{eq:t_fric_definition}
    t_\mathrm{fric}\equiv \frac{mv}{F_\mathrm{D}},
\end{equation}
with the particle mass $m$ given by
\begin{equation}
\label{eq:mass}
    m=\frac{4\pi}{3}\rho_\bullet s^3,
\end{equation}
with $\rho_\bullet$ the material density. Using this and Eq. \eqref{eq:epstein}, the friction time becomes
\begin{equation}
\label{eq:t_fric}
    t_\mathrm{fric}=\frac{\rho_\bullet}{\rho_\mathrm{gas}}\frac{s}{v_\mathrm{th}}.
\end{equation}
Particles are considered tightly bound to the gas if their Stokes number $\tau_\mathrm{fric}$, defined as the friction time times the local Kepler frequency, is smaller than 1/2 (see Appendix \ref{ap:settling} for the origin of this definition), as follows:
\begin{equation}
\label{eq:tau_fric}
    \tau_\mathrm{fric}\equiv t_\mathrm{fric}\Omega_\mathrm{K}<1/2,
\end{equation}
with $\Omega_\mathrm{K}$ the Kepler frequency. From Eq. \eqref{eq:t_fric_definition} we find this to be the case for particles with size
\begin{equation}
\label{eq:s_b}
    s<s_\mathrm{b}\equiv\frac{1}{2}\frac{\rho_\mathrm{gas}}{\rho_\bullet}\frac{v_\mathrm{th}}{\Omega_\mathrm{K}}.
\end{equation}
Although they are mainly moving along with the gas, these small particles are being pulled down by gravity, causing them to settle into the midplane. According to \citet{armitage}, the settling occurs at a  settling velocity found from equating the vertical component of gravity to the drag from this vertical motion, namely:\ 
\begin{equation}
    \frac{GM_*m}{R^2}\frac{z}{R}=\frac{4\pi}{3}\rho_\mathrm{gas}s^2v_\mathrm{th}v_\mathrm{settle}.
\end{equation}
Using Eq. \eqref{eq:mass} for the mass of a particle and substituting the Kepler frequency:
\begin{equation}
    \Omega_\mathrm{K}=\sqrt{\frac{GM_*}{R^3}},
\end{equation}
for the settling velocity, we find:
\algebra{
    \frac{GM_*}{R^2}\frac{z}{R}\frac{4\pi}{3}\rho_\bullet s^3&=\frac{4\pi}{3}\rho_\mathrm{gas}s^2v_\mathrm{th}v\\
    \frac{GM_*}{R^3}z\rho_\bullet s^3&=\rho_\mathrm{gas}s^2v_\mathrm{th}v\\
    \Omega_\mathrm{K}^2z\rho_\bullet s^3&=\rho_\mathrm{gas}s^2v_\mathrm{th}v\\
    \Omega_\mathrm{K}^2z\frac{\rho_\bullet}{\rho_\mathrm{gas}} \frac{s}{v_\mathrm{th}}&=v\\
}
\begin{equation}
    v_\mathrm{settle}=\frac{\rho_\bullet}{\rho_\mathrm{gas}}\frac{\Omega_\mathrm{K}^2}{v_\mathrm{th}}sz.
\end{equation}
On this basis, the timescale at which a particle settles into the midplane becomes:
\begin{align}
    t_\mathrm{settle}&\equiv\frac{z}{v_\mathrm{settle}}\\
    &=\frac{\rho_\mathrm{gas}}{\rho_\bullet}\frac{v_\mathrm{th}}{\Omega_\mathrm{K}^2}s^{-1}.
    \label{eq:t_settle}
\end{align}
This is what happens to small particles which are tightly bound to the gas. Larger particles do not leave their inclined orbits all together, but friction causes the inclination to be damped over the course of multiple orbits. The timescale on which these particles are pushed into the midplane is twice the friction time (see Appendix \ref{ap:settling} for the derivation), so the damping time is given by
\begin{equation}
    t_\mathrm{damp}=2\frac{\rho_\bullet}{\rho_\mathrm{gas}}\frac{s}{v_\mathrm{th}}.
    \label{eq:t_damp}
\end{equation}
\subsection{Collisional lifetime}
As described by \citet{vega_phenomenon}, the collisional lifetime, $t_\mathrm{coll}$, of a single particle in the collisional cascade is given by the volume $V$, in which the particles are located, divided by the volume sweeped by the particle per unit time:
\begin{equation}
    t_\mathrm{coll}(s)=\frac{V}{\nu v_\mathrm{K}\sigma_\mathrm{tot}(s)},
\end{equation}
with $\nu v_\mathrm{K}$ as the relative velocity between particles in terms of the Kepler velocity, $v_\mathrm{K}$, and $\sigma_\mathrm{tot}$ the total collisional cross-section. Below, we make the implicit assumption that the relative velocities result from inclinations of the orbits.  In reality, a similar component of relative velocities stems from the eccentricities of orbits. However, the excitation and damping factors affect both inclinations and eccentricities in similar ways, so we assume that the inclinations are a good proxy for the source of relative velocities. If the smallest particle capable of destroying a particle of size $s$ has a size of $\epsilon s$, the total collisional cross-section of a particle is given by:
\begin{align}
    \sigma_\mathrm{tot}(s)&=\int_{\epsilon s}^{s_\mathrm{pt}}{\pi\left(s+s'\right)^2f_\mathrm{cc}(s')\mathrm{d}s'},
\end{align}
with $f_\mathrm{cc}(s)$ the size distribution of particles in a steady-state collisional cascade and $s_\mathrm{pt}$ the size of the largest planetesimal in the cascade. Then, $\epsilon$ is given by \citet{vega_phenomenon} as:
\begin{equation}
    \epsilon^3=\frac{(\nu v_\mathrm{K})^2}{4S}-\sqrt{\frac{(\nu v_\mathrm{K})^4}{16S^2} - \frac{(\nu v_\mathrm{K})^2}{2S}}-1
,\end{equation}
with $S=200$ J kg$^{-1}$, a value close to the one suggested for practical use by \citet{2008ARA&A..46..339W}. As shown by \citet{dohnanyi} and later elaborated upon by \citet{tanaka_1996_steadystate}, $f_\mathrm{cc}(s)$ is given by
\begin{equation}
\label{eq:f_cc}
    f_\mathrm{cc}(s)=f_{0,\mathrm{cc}}s^{\gamma_\mathrm{cc}},
\end{equation}
with $f_{0,\mathrm{cc}}$ a constant and $\gamma_\mathrm{cc}=-3.5$. Since $\gamma_\mathrm{cc}<-3$ the cross-section is dominated by the smallest grains, and the integral above is evaluated as:
\begin{equation}
\label{eq:sigma_tot}
    \sigma_\mathrm{tot}(s)\approx \epsilon_0 f_{0,\mathrm{cc}}s^{\gamma_\mathrm{cc}+3},
\end{equation}
with
\begin{equation}
\label{eq:epsilon_0}
    \epsilon_0\equiv-\pi\epsilon^{\gamma_\mathrm{cc}+1}/(\gamma_\mathrm{cc}+1).
\end{equation}
Now, by substituting this into Eq. \eqref{eq:t_coll} and using $\gamma_\mathrm{cc}=-3.5,$ we find the collisional lifetime of a particle to be:
\begin{equation}
\label{eq:t_coll}
    t_\mathrm{coll}(s)=\frac{V}{\nu v_\mathrm{K}\epsilon_0}\frac{1}{f_{0,\mathrm{cc}}}s^{0.5}.
\end{equation}

\subsection{Removal from the cascade}
Due to the effects of gas described above, particles are removed from the collisional cascade. First of all, particles which are tightly coupled to the gas (with $s<s_\mathrm{b}$) move with the gas, removing the relative velocities between them, preventing them from colliding and fragmenting. Secondly, not well coupled particles are removed from the cascade if the time it takes for them to be damped into the midplane is shorter than the timescale at which they collide. Using Eqs. \eqref{eq:t_damp} and \eqref{eq:t_coll} we find that this is the case for particles with: 
\algebra{
    t_\mathrm{damp}&=t_\mathrm{coll}\\
    2t_\mathrm{fric}&=t_\mathrm{coll}\\
    2\frac{\rho_\bullet}{\rho_\mathrm{gas}}\frac{s}{v_\mathrm{th}}&=\frac{V}{\nu v_\mathrm{K}\epsilon_0}\frac{1}{f_{\mathrm{cc}}(s)s^3}\\
    2\frac{\rho_\bullet}{\rho_\mathrm{gas}}\frac{s}{v_\mathrm{th}}&=\frac{V}{\nu v_\mathrm{K}\epsilon_0}\frac{1}{f_{\mathrm{0,cc}}s^{-0.5}}\\
    2\frac{\rho_\bullet}{\rho_\mathrm{gas}}\frac{s^{0.5}}{v_\mathrm{th}}&=\frac{V}{\nu v_\mathrm{K}\epsilon_0}\frac{1}{f_{\mathrm{0,cc}}}\\
    s^{0.5}&=\frac{1}{2}\frac{\rho_\mathrm{gas}}{\rho_\bullet}\frac{V}{\nu v_\mathrm{K}\epsilon_0}\frac{v_\mathrm{th}}{f_{\mathrm{0,cc}}}\\
    }
\begin{equation}
\label{eq:s_cc}
    s<s_\mathrm{cc}\equiv\left(\frac{1}{2}\frac{\rho_\mathrm{gas}}{\rho_\bullet}\frac{V}{\nu v_\mathrm{K}\epsilon_0}\frac{v_\mathrm{th}}{f_{0,\mathrm{cc}}}\right)^2.
\end{equation}
In general, $s_\mathrm{cc}>s_\mathrm{b}$; therefore, all particles
larger than $s_\mathrm{cc}$ are involved in the collisional cascade
and produce smaller particles through fragmentation. The abundance of
small particles is what determines the cloud's optical depth, since
for power-laws with $\gamma$ steeper than $-3$, these dominate the
opacity.

\section{The broken cascade}
\label{sec:broken_cascade}
\subsection{Size distribution of small particles}
In the presence of gas, we have now a broken cascade where particles below a certain size no longer collide and fragment. These particles are still produced by the fragmentation of larger particles, but settling is what removes them from the higher regions of the dust cloud, so that they will no longer contribute to the optical depth of the cloud. Taking into account the production and removal of particles, the size distribution of bound particles $f_\mathrm{b}(s)$ is given by
\begin{equation}
\label{eq:size_dist_ODE}
    \frac{\mathrm{d}f_\mathrm{b}(s)}{\mathrm{d}t}=\phi_\mathrm{prod}(s)-\frac{f_\mathrm{b}(s)}{t_\mathrm{settle}(s)},
\end{equation}
with $\phi_\mathrm{prod}(s)\mathrm{d}s$ the rate at which particles with $s\in[s,s+\mathrm{d}s]$ are being produced. The steady-state solution to this equation (see appendix \ref{ap:ODE} for the complete solution) is
\begin{equation}
\label{eq:f_b_steady_state}
    f_\mathrm{b}(s)=t_\mathrm{settle}(s)\phi_\mathrm{prod}(s),
\end{equation}
which is reached after a couple of settling timescales.

\subsection{Production rate of small particles}
The production rate of these small ($s\ll s_\mathrm{b}$) particles can be derived by considering the result of the fragmentation of a single large, unbound particle. As shown by  \citet{Takasawa_2011} and \citet{OBrien}, for instance, the fragments resulting from high-velocity collisions (such as those in our system) can be assumed to follow a power-law size distribution, namely:
\begin{equation}
\label{eq:f_frag}
    f_\mathrm{frag}(s,s_0)=f_{0,\mathrm{frag}}(s_0)s^{\gamma_\mathrm{f}},
\end{equation}
with $s_0$ the initial particle's size, and $\gamma_\mathrm{f}$ between $-3$ and $-6$. Since we are mainly interested in the production of small particles, a more negative value for this power-law would be beneficial. For now we assume the worst case scenario where $\gamma_\mathrm{f}=-3$. Using that, assuming the largest fragment to be half the size of the initial particle, mass-conservation requires \citep{OBrien}:
\begin{align}
    m=\frac{4\pi}{3}\rho_\bullet  s_0^3&=\int_{s_\mathrm{min}}^{s_0/2}{\frac{4\pi}{3}\rho_\bullet s^3f_\mathrm{frag}(s,s_0)\mathrm{d}s}\\
    \label{eq:fragmentation_integral}
    &=\frac{4\pi}{3}\rho_\bullet\frac{1}{\gamma_\mathrm{f}+4}\left[\left(\frac{s_0}{2}\right)^{\gamma_\mathrm{f}+4}-s_\mathrm{min}^{\gamma_\mathrm{f}+4}\right]f_{0,\mathrm{frag}}.
\end{align}
Since for $\gamma_\mathrm{f}>-4$ the large fragments dominate the mass, solving for $f_{0,\mathrm{frag}}$ yields
\begin{equation}
\label{eq:f_0frag}
    f_{0,\mathrm{frag}}(s_0)=(\gamma_\mathrm{f}+4)2^{\gamma_\mathrm{f}+4}s_0^{-\gamma_\mathrm{f}-1}.
\end{equation}
Particles smaller than $s_\mathrm{cc}/2$ are produced by all particles in the collisional cascade: from $s_\mathrm{cc}$ up to the largest planetesimals of size $s_\mathrm{pt}$, and the total production rate of particles with sizes of $\in[s,s+\mathrm{d}s]$ is given by:
\begin{equation}
\label{eq:phi_prod_integral}
    \phi_\mathrm{prod}(s)=\int_{s_\mathrm{cc}}^{s_\mathrm{pt}}{\frac{f_\mathrm{cc}(s_0)}{t_\mathrm{coll}(s_0)}f_\mathrm{frag}(s,s_0)\mathrm{d}s_0}.
\end{equation}
Using Eq. \eqref{eq:t_coll} for the collision time $t_\mathrm{coll}$, and Eq. \eqref{eq:f_cc} for the size distribution of particles in the collisional cascade $f_\mathrm{cc}$, we find:
\algebra{
    \phi_\mathrm{prod}(s)&=\int_{s_\mathrm{cc}}^{s_\mathrm{pt}}{\frac{f_{0,\mathrm{cc}}s_0^{-3.5}}{\frac{V}{\nu v_\mathrm{K}\epsilon_0}\frac{1}{f_{\mathrm{cc}}(s_0)s_0^3}}f_\mathrm{frag}(s,s_0)\mathrm{d}s_0}\\
    &=\int_{s_\mathrm{cc}}^{s_\mathrm{pt}}{\frac{f_{0,\mathrm{cc}}s_0^{-3.5}}{\frac{V}{\nu v_\mathrm{K}\epsilon_0}\frac{1}{f_{0,\mathrm{cc}}s_0^{-3.5}s_0^3}}f_\mathrm{frag}(s,s_0)\mathrm{d}s_0}\\
    &=\int_{s_\mathrm{cc}}^{s_\mathrm{pt}}{\frac{f_{0,\mathrm{cc}}^2s_0^{-4}}{\frac{V}{\nu v_\mathrm{K}\epsilon_0}}f_\mathrm{frag}(s,s_0)\mathrm{d}s_0}\\
    &=\frac{\nu v_\mathrm{K}\epsilon_0}{V}f_{0,\mathrm{cc}}^2\int_{s_\mathrm{cc}}^{s_\mathrm{pt}}{s_0^{-4}f_\mathrm{frag}(s,s_0)\mathrm{d}s_0}.
}
\begin{equation}
\label{eq:phi_prod_general}
    \phi_\mathrm{prod}(s)=\frac{\nu v_\mathrm{K}\epsilon_0}{V}f_{0,\mathrm{cc}}^2\int_{s_\mathrm{cc}}^{s_\mathrm{pt}}{s_0^{-4}f_\mathrm{frag}(s,s_0)\mathrm{d}s_0}.
\end{equation}
In the case where $\gamma_\mathrm{f}>-4$, substitution of Eq. \eqref{eq:f_frag} for $f_{0,\mathrm{frag}}$ and using $s_\mathrm{pt}\gg s_\mathrm{cc}$ yields
\algebra{
    \phi_\mathrm{prod}(s)&=\frac{\nu v_\mathrm{K}\epsilon_0}{V}f_{0,\mathrm{cc}}^2\int_{s_\mathrm{cc}}^{s_\mathrm{pt}}{s_0^{-4}(\gamma_\mathrm{f}+4)2^{\gamma_\mathrm{f}+4}s_0^{-\gamma_\mathrm{f}-1}s^\gamma_\mathrm{f}\mathrm{d}s_0}\\
    &=\frac{\nu v_\mathrm{K}\epsilon_0}{V}f_{0,\mathrm{cc}}^2(\gamma_\mathrm{f}+4)2^{\gamma_\mathrm{f}+4}s^\gamma_\mathrm{f}\int_{s_\mathrm{cc}}^{s_\mathrm{pt}}{s_0^{-4}s_0^{-\gamma_\mathrm{f}-1}\mathrm{d}s_0}\\
    &=\frac{\nu v_\mathrm{K}\epsilon_0}{V}f_{0,\mathrm{cc}}^2(\gamma_\mathrm{f}+4)2^{\gamma_\mathrm{f}+4}s^\gamma_\mathrm{f}\int_{s_\mathrm{cc}}^{s_\mathrm{pt}}{s_0^{-\gamma_\mathrm{f}-5}\mathrm{d}s_0}\\
    &=\frac{\nu v_\mathrm{K}\epsilon_0}{V}f_{0,\mathrm{cc}}^2(\gamma_\mathrm{f}+4)2^{\gamma_\mathrm{f}+4}s^\gamma_\mathrm{f}\frac{1}{\gamma_\mathrm{f}+4}\left(s_\mathrm{pt}^{-\gamma_\mathrm{f}-4}-s_\mathrm{cc}^{-\gamma_\mathrm{f}-4} \right)
}
\begin{align}
    \phi_\mathrm{prod}(s)
    &=2^{\gamma_\mathrm{f}+4}\frac{\nu v_\mathrm{K}\epsilon_0}{V}f_{0,\mathrm{cc}}^2 \left(s_\mathrm{cc}^{-\gamma_\mathrm{f}-4}-s_\mathrm{pt}^{-\gamma_\mathrm{f}-4} \right)s^{\gamma_\mathrm{f}}\\
    &\approx 2^{\gamma_\mathrm{f}+4}\frac{\nu v_\mathrm{K}\epsilon_0}{V}f_{0,\mathrm{cc}}^2s_\mathrm{cc}^{-\gamma_\mathrm{f}-4}s^{\gamma_\mathrm{f}}
    \label{eq:phi_prod_large_gamma}.
\end{align}
Using $\gamma_\mathrm{f}=-3$ and substituting Eq. \eqref{eq:s_cc} for $s_\mathrm{cc}$, we find:
\algebra{
    \phi_\mathrm{prod}(s)
    &=2\frac{\nu v_\mathrm{K}\epsilon_0}{V}f_{0,\mathrm{cc}}^2s_\mathrm{cc}^{-1} s^{-3}\\
    &=2\frac{\nu v_\mathrm{K}\epsilon_0}{V}f_{0,\mathrm{cc}}^2\left(\frac{1}{2}\frac{\rho_\mathrm{gas}}{\rho_\bullet}\frac{V}{\nu v_\mathrm{K}\epsilon_0}\frac{v_\mathrm{th}}{f_{0,\mathrm{cc}}}\right)^{-2} s^{-3}\\
    &=2\frac{\nu v_\mathrm{K}\epsilon_0}{V}f_{0,\mathrm{cc}}^2\left(2\frac{\rho_\bullet}{\rho_\mathrm{gas}}\frac{\nu v_\mathrm{K}\epsilon_0}{V}\frac{f_{0,\mathrm{cc}}}{v_\mathrm{th}}\right)^{2} s^{-3}\\
    &=2 \frac{\nu v_\mathrm{K}\epsilon_0}{V}f_{0,\mathrm{cc}}^2\left(2\frac{\rho_\bullet}{\rho_\mathrm{gas}}\frac{\nu v_\mathrm{K}\epsilon_0}{V}\frac{f_{0,\mathrm{cc}}}{v_\mathrm{th}}\right)^{2} s^{-3}\\
    &=8\left(\frac{\rho_\bullet}{\rho_\mathrm{gas}}\right)^2\frac{\nu v_\mathrm{K}\epsilon_0}{V}f_{0,\mathrm{cc}}^2\left(\frac{\nu v_\mathrm{K}\epsilon_0}{V}\frac{f_{0,\mathrm{cc}}}{v_\mathrm{th}}\right)^{2} s^{-3}\\
    &=8\left(\frac{\rho_\bullet}{\rho_\mathrm{gas}}\right)^2\left(\frac{\nu v_\mathrm{K}\epsilon_0}{V}\right)^{3}f_{0,\mathrm{cc}}^2\left(\frac{f_{0,\mathrm{cc}}}{v_\mathrm{th}}\right)^{2} s^{-3}\\
    &=8\left(\frac{\rho_\bullet}{\rho_\mathrm{gas}}\right)^2\left(\frac{\nu v_\mathrm{K}\epsilon_0}{V}\right)^{3}f_{0,\mathrm{cc}}^4\left(\frac{1}{v_\mathrm{th}}\right)^{2} s^{-3}\\
    &=8\left(\frac{\rho_\bullet}{\rho_\mathrm{gas}}\right)^2\frac{1}{v_\mathrm{th}^2}\left(\frac{\nu v_\mathrm{K}\epsilon_0}{V}\right)^{3}f_{0,\mathrm{cc}}^4 s^{-3}\\
}
\begin{equation}
\label{eq:phi_prod2}
    \phi_\mathrm{prod}(s)=\frac{8}{v_\mathrm{th}^2}\left(\frac{\rho_\bullet}{\rho_\mathrm{gas}}\right)^2\left(\frac{\nu v_\mathrm{K}\epsilon_0}{V}\right)^{3}f_{0,\mathrm{cc}}^4 s^{-3}.
\end{equation}

\subsection{Replenishment of large particles}
As we can see, the amount of small particles is dependent on the abundance of large, uncoupled particles through $f_{0,\mathrm{cc}}$. To compensate for the constant destruction of these uncoupled particles, we assume that the largest particles in the cascade are constantly being replenished by planetesimals being injected into the system. This can happen in two ways. If there is a population of somewhat larger planetesimals with collisional lifetimes equal to or larger than the lifetime of the disk phase we are talking about (typically less that 10\,Myr), the mass injection into the cascade could simply come from collisions of these planetesimals, but would require a sufficient amount of these larger bodies to be present \citep{2008ApJ...673.1123L,Krivov_Wyatt_2020}. These planetesimals would have to be collisionally excited by planets early on in the disk.  An alternative is that smaller planetesimals with shorter collisional lifetimes are injected continuously into relevant orbits by interactions with planets in the disk, for example, by injecting bodies from further out in the disk by a planet there \citep{Raymond_Izidoro_2017} or simply by scattering a local planetesimal into a high-inclination orbit \citep{2011A&A...531A..80K}.

Assuming the size of the largest fragment resulting from the destruction of a particle to be half of that of the initial particle, the mass flux required to keep up the population of the largest particles with $s\in[s_\mathrm{pt}/2,s_\mathrm{pt}]$, is:
\begin{equation}
\label{eq:M_in_integral}
    \dot{M}_\mathrm{in}=\int_{s_\mathrm{pt}/2}^{s_\mathrm{pt}}{\frac{m(s)}{t_\mathrm{coll}(s)}f_\mathrm{cc}(s)\mathrm{d}s}.
\end{equation}
Using Eq. \eqref{eq:mass} for $m$ and Eq. \eqref{eq:t_coll} for $t_\mathrm{coll}$,  for the mass flux we find:\algebra{
    \d\ot{M}_\mathrm{in}&=\int_{s_\mathrm{pt}/2}^{s_\mathrm{pt}}{\frac{\frac{4\pi}{3}\rho_\bullet s^3}{\frac{V}{\nu v_\mathrm{K}\epsilon_0}\frac{1}{f_{\mathrm{cc}}(s)s^3}}f_\mathrm{cc}(s)\mathrm{d}s}\\
    &=\frac{4\pi}{3}\rho_\bullet\frac{\nu v_\mathrm{K}\epsilon_0}{V}\int_{s_\mathrm{pt}/2}^{s_\mathrm{pt}}{s^6f_\mathrm{cc}(s)^2\mathrm{d}s}\\
    &=\frac{4\pi}{3}\rho_\bullet\frac{\nu v_\mathrm{K}\epsilon_0}{V}f_{0,\mathrm{cc}}^2\int_{s_\mathrm{pt}/2}^{s_\mathrm{pt}}{s^{-1}\mathrm{d}s}\\
}
\begin{equation}
\label{eq:M_in_from_f_cc}
    \dot{M}_\mathrm{in}=\frac{4\pi}{3}\rho_\bullet\frac{\nu v_\mathrm{K}\epsilon_0}{V}f_{0,\mathrm{cc}}^2\ln{2}.
\end{equation}
Solving for $f_{0,\mathrm{cc}}$ we can express the abundance of (uncoupled) particles in the collisional cascade in terms of the cloud's dimensions and the mass flux, namely, 
\algebra{
    \frac{\nu v_\mathrm{K}\epsilon_0}{V}f_{0,\mathrm{cc}}^2=\frac{1}{\ln{2}}\frac{\dot{M}_\mathrm{in}}{\frac{4\pi}{3}\rho_\bullet}
}
\begin{equation}
\label{eq:f_0cc}
    f_{0,\mathrm{cc}}=\sqrt{\frac{1}{\ln{2}}\frac{V}{\nu v_\mathrm{K}\epsilon_0}\frac{\dot{M}_\mathrm{in}}{\frac{4\pi}{3}\rho_\bullet}}.
\end{equation}
Substituting this for $f_{0,\mathrm{cc}}$ into Eq. \eqref{eq:phi_prod2}, we find the production rate of small particles to be:
\algebra{
    \phi_\mathrm{prod}(s)
    &=8\left(\frac{\rho_\bullet}{\rho_\mathrm{gas}}\right)^2\frac{1}{v_\mathrm{th}^2}\left(\frac{\nu v_\mathrm{K}\epsilon_0}{V}\right)^{3}f_{0,\mathrm{cc}}^4 s^{-3}\\
    &=8\left(\frac{\rho_\bullet}{\rho_\mathrm{gas}}\right)^2\frac{1}{v_\mathrm{th}^2}\left(\frac{\nu v_\mathrm{K}\epsilon_0}{V}\right)^{3}\left(\frac{1}{\ln{2}}\frac{V}{\nu v_\mathrm{K}\epsilon_0}\frac{\dot{M}_\mathrm{in}}{\frac{4\pi}{3}\rho_\bullet}\right)^2 s^{-3}\\
    &=8\frac{1}{\rho_\mathrm{gas}^2}\frac{1}{v_\mathrm{th}^2}\frac{\nu v_\mathrm{K}\epsilon_0}{V}\left(\frac{1}{\ln{2}}\frac{\dot{M}_\mathrm{in}}{\frac{4\pi}{3}}\right)^2 s^{-3}\\
    &=\frac{9}{2\pi^2(\ln{2})^2}\frac{1}{\rho_\mathrm{gas}^2v_\mathrm{th}^2}\frac{\nu v_\mathrm{K}\epsilon_0}{V}\dot{M}_\mathrm{in}^2 s^{-3}\\
}
\begin{equation}
\label{eq:phi_prod3}
        \phi_\mathrm{prod}(s)=\frac{9}{2\pi^2(\ln{2})^2}\frac{1}{\rho_\mathrm{gas}^2v_\mathrm{th}^2}\frac{\nu v_\mathrm{K}\epsilon_0}{V}\dot{M}_\mathrm{in}^2 s^{-3}
.\end{equation}
\subsection{The broken distribution}
We can now find the distribution of coupled particles by substituting the equation for the production rate above and Eq. \eqref{eq:t_settle} into Eq. \eqref{eq:f_b_steady_state} for $f_\mathrm{b}$, which yields:
\algebra{
    f_\mathrm{b}(s)
    &=t_\mathrm{settle}(s)\phi_\mathrm{prod}(s)\\
    &=\frac{\rho_\mathrm{gas}}{\rho_\bullet}\frac{v_\mathrm{th}}{\Omega_\mathrm{K}^2}s^{-1} \cdot \frac{9}{2\pi^2(\ln{2})^2}\frac{1}{\rho_\mathrm{gas}^2v_\mathrm{th}^2}\frac{\nu v_\mathrm{K}\epsilon_0}{V}\dot{M}_\mathrm{in}^2 s^{-3}\\
    &=\frac{9}{2\pi^2(\ln{2})^2}\frac{\rho_\mathrm{gas}}{\rho_\bullet}\frac{v_\mathrm{th}}{\Omega_\mathrm{K}^2} \frac{1}{\rho_\mathrm{gas}^2v_\mathrm{th}^2}\frac{\nu v_\mathrm{K}\epsilon_0}{V}\dot{M}_\mathrm{in}^2 s^{-4}\\
    &=\frac{9}{2\pi^2(\ln{2})^2}\frac{1}{\Omega_\mathrm{K}^2} \frac{1}{\rho_\mathrm{gas}\rho_\bullet v_\mathrm{th}}\frac{\nu v_\mathrm{K}\epsilon_0}{V}\dot{M}_\mathrm{in}^2 s^{-4}\\
}
\begin{equation}
\label{eq:f_b}
    f_\mathrm{b}(s)=\frac{9}{2\pi^2(\ln{2})^2}\frac{1}{\Omega_\mathrm{K}^2} \frac{1}{\rho_\mathrm{gas}\rho_\bullet v_\mathrm{th}}\frac{\nu v_\mathrm{K}\epsilon_0}{V}\dot{M}_\mathrm{in}^2 s^{-4}.
\end{equation}
We can paint a complete picture of the population of dust particles in the cloud by applying same logic to particles which are not tightly coupled to the gas, but removed from the cascade through damping ($s_\mathrm{b}<s<s_\mathrm{cc}$). By using $t_\mathrm{damp}$ instead of $t_\mathrm{settle}$ we find for these intermediate-size particles
\algebra{
    f_\mathrm{x}(s)
    &=2\frac{\rho_\bullet}{\rho_\mathrm{gas}}\frac{s}{v_\mathrm{th}}\frac{9}{2\pi^2(\ln{2})^2}\frac{1}{\rho_\mathrm{gas}^2v_\mathrm{th}^2}\frac{\nu v_\mathrm{K}\epsilon_0}{V}\dot{M}_\mathrm{in}^2 s^{-3}\\
    &=\frac{9}{\pi^2(\ln{2})^2}\rho_\bullet s\frac{1}{\rho_\mathrm{gas}^3v_\mathrm{th}^3}\frac{\nu v_\mathrm{K}\epsilon_0}{V}\dot{M}_\mathrm{in}^2 s^{-3}\\
    &=\frac{9}{\pi^2(\ln{2})^2}\frac{\rho_\bullet}{\rho_\mathrm{gas}^3v_\mathrm{th}^3}\frac{\nu v_\mathrm{K}\epsilon_0}{V}\dot{M}_\mathrm{in}^2 s^{-2}\\
}
\begin{equation}
\label{eq:f_int}
    f_\mathrm{int}(s)=\frac{9}{\pi^2(\ln{2})^2}\frac{\rho_\bullet}{\rho_\mathrm{gas}^3v_\mathrm{th}^3}\frac{\nu v_\mathrm{K}\epsilon_0}{V}\dot{M}_\mathrm{in}^2 s^{-2} .
\end{equation}
We can now compare two cases: first of all, in the complete absence of gas, a normal collisional cascade emerges in which planetesimals of size, $s_\mathrm{pt}$, are supplied and particles below $s_\mathrm{min}$ are removed (by radiation pressure, stellar wind or Poynting-Robertson drag \citep{Lagrange-PPVI,2008ARA&A..46..339W}). The complete size distribution is
\begin{equation}
    f(s)=
    \begin{cases}
        f_{0,\mathrm{cc}}s^{-3.5} & s_\mathrm{min}<s<s_\mathrm{pt}\\
        0 & \mathrm{otherwise},
    \end{cases}
\end{equation}
with $f_{0,\mathrm{cc}}$ given by Eq. \eqref{eq:f_0cc}.

Secondly, in the presence of gas, particles smaller than $s_\mathrm{b}$ (given by Eq. \eqref{eq:s_b}) are tightly coupled to the gas and follow a distribution with a steeper slope given by Eq. \eqref{eq:f_b}. Furthermore, intermediate size particles smaller than $s_\mathrm{cc}$ have a reduced abundance, since they are damped into the midplane quicker than they collide, resulting in the distribution given by Eq. \eqref{eq:f_int}. The complete size distribution for this case is:
\begin{equation}
    f(s)=
    \begin{cases}
        f_{0,\mathrm{cc}}s^{-3.5} & s_\mathrm{cc}<s<s_\mathrm{pt} \\
        f_{0,\mathrm{int}}s^{-2} & s_\mathrm{b}<s<s_\mathrm{cc} \\
        f_{0,\mathrm{b}}s^{-4} & s_\mathrm{min}<s<s_\mathrm{b} \\
        0 & \mathrm{otherwise},
    \end{cases}
\end{equation}
with $f_{0,\mathrm{int}}$ given by
\begin{equation}
\label{eq:f_0int}
    f_{0,\mathrm{int}}=\frac{9}{\pi^2(\ln{2})^2}\frac{\rho_\bullet}{\rho_\mathrm{gas}^3v_\mathrm{th}^3}\frac{\nu v_\mathrm{K}\epsilon_0}{V}\dot{M}_\mathrm{in}^2
\end{equation} 
and $f_{0,\mathrm{b}}$ by
\begin{equation}
\label{eq:f_0b}
    f_{0,\mathrm{b}}=\frac{9}{2\pi^2(\ln{2})^2}\frac{1}{\Omega_\mathrm{K}^2} \frac{1}{\rho_\mathrm{gas}\rho_\bullet v_\mathrm{th}}\frac{\nu v_\mathrm{K}\epsilon_0}{V}\dot{M}_\mathrm{in}^2.
  \end{equation}

\section{Required mass flux}
\label{sec:required_mass_flux}
\subsection{An opaque cloud}
As we have shown above, both the amount of particles in the collisional cascade as well as the amount of coupled particles is determined by the mass flux $\dot{M}_\textrm{in}$. To explain observations, a cloud of particles is needed that is more or less opaque to stellar (optical) radiation. This requires a certain minimum amount of particles and, consequentially, requires a certain mass flux. The optical depth of a cloud of dust particles (see Fig. \ref{fig:system} for the dimensions) following a size distribution given by $f(s)$ between $s_\textrm{min}$ and $s_\textrm{max}$ is given by
\begin{equation}
\label{eq:opacity_integral}
    \tau_\textrm{g}=\int_{s_\textrm{min}}^{s_\textrm{max}}\frac{\sigma_\textrm{g}(s)}{V/\Delta R}f(s)\textrm{d}s,
\end{equation}
where we use, for simplicity, the geometric cross section $\sigma_\textrm{g}$ for the absorption of light, given by
\begin{equation}
\label{eq:sigma_g}
    \sigma_\textrm{g}=\pi s^2.
\end{equation}
Using this cross-section and a general power-law $f(s)=f_0s^\gamma$, integration yields:
\begin{equation}
    \tau_\textrm{g}=\frac{\pi \Delta R}{V}f_0\frac{s_\textrm{max}^{\gamma+3}-s_\textrm{min}^{\gamma+3}}{\gamma+3}.
\end{equation}
By solving for $f_0$ we find the minimum amount of particles needed for a given geometric optical depth as:
\begin{equation}
\label{eq:f_0_min_complete}
    f_{0,\mathrm{min}}=\frac{V}{\pi\Delta R}\frac{\gamma+3}{s_\textrm{max}^{\gamma+3}-s_\textrm{min}^{\gamma+3}}\tau_\textrm{g}.
\end{equation}
If $\gamma<-3$ (which is the case for small particles in gas as well as for particles in the collisional cascade, see Eqs. \ref{eq:f_cc} and \ref{eq:f_b}) the optical depth is dominated by small particles, and this can be approximated as:
\begin{equation}
\label{eq:f_0min}
    f_{0,\mathrm{min}}=\frac{V}{\pi\Delta R}\frac{-(\gamma+3)}{s_\textrm{min}^{\gamma+3}}\tau_\textrm{g}.
\end{equation}
\subsection{Mass flux in the presence of gas}
By equating the amount of small particles $f_{0,\mathrm{b}}$ (Eq. \eqref{eq:f_0b}) to the minimum amount of particles required, found from Eq. \eqref{eq:f_0min} with $\gamma=-4$, we can find the mass flux required for a given optical depth in the case where there is gas in the system. This results in
\algebra{
    \frac{9}{2\pi^2(\ln{2})^2}\frac{1}{\Omega_\mathrm{K}^2} \frac{1}{\rho_\mathrm{gas}\rho_\bullet v_\mathrm{th}}\frac{\nu v_\mathrm{K}\epsilon_0}{V}\dot{M}_\mathrm{in}^2 &= \frac{V}{\pi\Delta R}\frac{-(\gamma+3)}{s_\textrm{min}^{\gamma+3}}\tau_\textrm{g}\\
    \frac{9}{2\pi^2(\ln{2})^2}\frac{1}{\Omega_\mathrm{K}^2} \frac{1}{\rho_\mathrm{gas}\rho_\bullet v_\mathrm{th}}\frac{\nu v_\mathrm{K}\epsilon_0}{V}\dot{M}_\mathrm{in}^2 &= \frac{V}{\pi\Delta R}s_\textrm{min}\tau_\textrm{g}\\
    \dot{M}_\mathrm{in}^2 &= \frac{2\pi^2(\ln{2})^2}{9}\Omega_\mathrm{K}^2 \rho_\mathrm{gas}\rho_\bullet v_\mathrm{th}\frac{V}{\nu v_\mathrm{K}\epsilon_0}\frac{V}{\pi\Delta R}s_\textrm{min}\tau_\textrm{g}\\
    \dot{M}_\mathrm{in} &= \frac{\sqrt{2}\pi\ln{2}}{3}\Omega_\mathrm{K}\sqrt{ \rho_\mathrm{gas}\rho_\bullet v_\mathrm{th}}\frac{V}{\sqrt{\nu v_\mathrm{K}\epsilon_0\pi\Delta R}}\sqrt{s_\textrm{min}\tau_\textrm{g}}\\
}
\begin{equation}
\label{eq:M_in_with_rough}
    \dot{M}_\mathrm{in}=\frac{\sqrt{2}\pi\ln{2}}{3}\Omega_\mathrm{K}V\sqrt{\frac{\rho_\mathrm{gas}\rho_\bullet v_\mathrm{th}s_\mathrm{min}\tau_\mathrm{g}}{\nu v_\mathrm{K}\epsilon_0\pi\Delta R}},
\end{equation}
which can be simplified by making some rough approximations for the cloud's structure. First of all, we assume the cloud to be a torus-like shell section of width $\Delta R$ at a distance $R$ to the central star, with an opening half angle $\alpha$ (see Fig. 1). In this way the cloud covers a fraction of $\eta \approx \tan{\alpha}$ of the total solid angle from the star. The volume can be approximated as:
\begin{equation}
\label{eq:volume}
    V=4\pi R^2 \Delta R \eta.
\end{equation}
Furthermore, we approximate the Kepler frequency $\Omega_\mathrm{K}$ (and consequentially the Kepler velocity with $v_\mathrm{K}=\Omega_\mathrm{K}R$) with:
\begin{equation}
    \Omega_\mathrm{K}=\sqrt{\frac{GM_*}{R^3}}
\end{equation}
for all particles. Furthermore, the cloud's height is approximately $H=\eta R$, and assuming homogeneity the gas density becomes:
\begin{equation}
\label{eq:rho_gas}
    \rho_\mathrm{gas}=\frac{\Sigma_\mathrm{gas}}{\eta R},
\end{equation}
with $\Sigma_\mathrm{gas}$ the surface density. Finally, given that the largest particles are on orbits with inclinations up to $\alpha$. Therefore, their vertical velocities can get as high as $v_\mathrm{K}\tan{\alpha}=\eta v_\mathrm{K}$. Since it is mainly these vertical velocity differences that will cause particles to collide, an appropriate number for the average relative velocity between particles is of this order of magnitude, which is why we use $\nu=\eta/4$ \citep[][for a typical inclination of 15 degrees]{Lissauer_Steward_PPIII}. Using all of this and Eq. \eqref{eq:v_th} for $v_\mathrm{th}$, we can evaluate Eq. \eqref{eq:M_in_with_rough} to:
\algebra{
    \dot{M}_\mathrm{in} = \frac{16}{3}\ln{2}\sqrt{\sqrt{\frac{2\pi^5GM_*k_\mathrm{B}T_\mathrm{gas}R}{\mu_\mathrm{mol}m_\mathrm{p}}}\frac{\Delta R\Sigma_\mathrm{gas}\rho_\bullet s_\mathrm{min}}{\epsilon_0}}
}
\begin{align}
\label{eq:mass_flux}
\dot{M}_\textrm{in}=
1.7\times10^{-4}\textrm{ }M_\oplus\textrm{yr}^{-1}
    \left(\frac{M_*}{M_\odot}\right)^{0.25}
    \left(\frac{R}{1\textrm{ au}}\right)^{0.25}
    \left(\frac{\Delta R}{0.1\textrm{ au}}\right)^{0.5}\nonumber\\
    \left(\frac{T_\mathrm{gas}}{300\textrm{ K}}\right)^{0.25}
    \left(\frac{\Sigma_\textrm{gas}}{1000 \textrm{ kg/m}^2}\right)^{0.5}
    \left(\frac{\rho_\bullet}{1500 \textrm{ kg/m}^3}\right)^{0.25}
    \left(\frac{s_\textrm{min}}{0.1 \textrm{ }\mu\textrm{m}}\right)^{0.5}
    \tau_\textrm{g}^{0.5}.
\end{align}

\subsection{Mass flux without gas}
The same can be done for the case of a classical debris disk where there is no dynamically relevant gas in the system, and all particles are in the collisional cascade. Equating $f_{0,\mathrm{min}}$ from Eq. \eqref{eq:f_0min} with $\gamma=-3.5$ to $f_{0,\mathrm{cc}}$ from Eq. \eqref{eq:f_0cc}, solving for the mass flux yields:
\algebra{
    \sqrt{\frac{1}{\ln{2}}\frac{V}{\nu v_\textrm{K}\epsilon_0}\frac{\dot{M}_\textrm{in}}{\frac{4\pi}{3}\rho_\bullet}}&=\frac{V}{2\pi\Delta R}s_\textrm{min}^{0.5}\tau_\textrm{g}\\
    \frac{1}{\ln{2}}\frac{V}{\nu v_\textrm{K}\epsilon_0}\frac{\dot{M}_\textrm{in}}{\frac{4\pi}{3}\rho_\bullet}&=\frac{V^2}{4\pi^2\Delta R^2}s_\textrm{min}\tau_\textrm{g}^2\\
    \dot{M}_\textrm{in}&=\ln{2}\frac{4\pi}{3}\rho_\bullet\frac{\nu v_\textrm{K}\epsilon_0}{V}\frac{V^2}{4\pi^2\Delta R^2}s_\textrm{min}\tau_\textrm{g}^2\\
    \dot{M}_\textrm{in}&=\frac{\ln{2}}{3\pi}\rho_\bullet\frac{\nu v_\textrm{K}\epsilon_0}{V}\frac{V^2}{\Delta R^2}s_\textrm{min}\tau_\textrm{g}^2\\
    \dot{M}_\textrm{in}&=\frac{\ln{2}}{3\pi}\rho_\bullet\nu v_\textrm{K}\epsilon_0\frac{V}{\Delta R^2}s_\textrm{min}\tau_\textrm{g}^2\\
}
\begin{equation}
    \dot{M}_\textrm{in}=\frac{\ln{2}}{3\pi}\rho_\bullet\nu v_\textrm{K}\epsilon_0\frac{V}{\Delta R^2}s_\textrm{min}\tau_\textrm{g}^2.
\end{equation}
Using the same approximations for $V$, $\nu$ and $v_\mathrm{K}$ as described above, this can be expressed as:
\algebra{
    \dot{M}_\textrm{in}&=\frac{\ln{2}}{12\pi}\rho_\bullet\eta v_\textrm{K}\epsilon_0\frac{V}{\Delta R^2}s_\textrm{min}\tau_\textrm{g}^2\\
    &=\frac{\ln{2}}{12\pi}\rho_\bullet\eta v_\textrm{K}\epsilon_0\frac{4\pi R^2 \Delta R\eta}{\Delta R^2}s_\textrm{min}\tau_\textrm{g}^2\\
    &=\frac{\ln{2}}{3}\rho_\bullet\eta v_\textrm{K}\epsilon_0\frac{R^2\eta}{\Delta R}s_\textrm{min}\tau_\textrm{g}^2\\
    &=\frac{\ln{2}}{3}\rho_\bullet\eta \sqrt{\frac{GM_*}{R}}\epsilon_0\frac{R^2\eta}{\Delta R}s_\textrm{min}\tau_\textrm{g}^2\\
    &=\frac{\ln{2}}{3}\rho_\bullet\eta \sqrt{GM_*}\epsilon_0\frac{R^{1.5}\eta}{\Delta R}s_\textrm{min}\tau_\textrm{g}^2\\
    &=\frac{\ln{2}}{3}\rho_\bullet\eta^2 G^{0.5}M_*^{0.5}\epsilon_0 R^{1.5}\Delta R^{-1}s_\textrm{min}\tau_\textrm{g}^2
}
\begin{align}
\label{eq:mass_flux_without}
\dot{M}_\textrm{in}=
2.4\times10^{-3}\textrm{ }M_\oplus/\textrm{yr }
    \left(\frac{M_*}{M_\odot}\right)^{0.5}
    \left(\frac{R}{1\textrm{ au}}\right)^{1.5}
    \left(\frac{\Delta R}{0.1\textrm{ au}}\right)^{-1}
    \left(\frac{\eta}{0.3}\right)^{2}\nonumber\\
    \left(\frac{\rho_\bullet}{1500 \textrm{ kg/m}^3}\right)
    \left(\frac{s_\textrm{min}}{0.1 \textrm{ }\mu\textrm{m}}\right)
    \tau_\textrm{g}^2.
\end{align}

\subsection{Influence of fragment size distribution}
\label{subsec:gamma_f}
Above, $\gamma_\mathrm{f}=-3$ is used for the size distribution of the particles resulting from a fragmentation event. This is a very conservative assumption, however, as \citet{Takasawa_2011} showed that $\gamma_\mathrm{f}$ can be as low as $-6$ for collisions at the velocities we are dealing with. As we are mainly interested in the production of small particles, lower values of $\gamma_\mathrm{f}$ would be beneficial and would bring down the required mass flux even further.
By solving the integrals in Eqs. \eqref{eq:fragmentation_integral},
\eqref{eq:phi_prod_integral} and \eqref{eq:opacity_integral}
numerically, we can find $f_{0,\mathrm{min}}$ and $f_{0,\mathrm{b}}$
for different values of $\gamma_\mathrm{f}$ as a function of
$\dot{M}_\mathrm{in}$. Root finding can be used to find the value of
$\dot{M}_\mathrm{in}$ for which $f_{0,\mathrm{b}}=f_{0,\mathrm{min}}$.

\section{Results and discussion}
\label{sec:results}
\subsection{Mass flux comparison}

\begin{figure*}
    \includegraphics[width=0.5\textwidth]{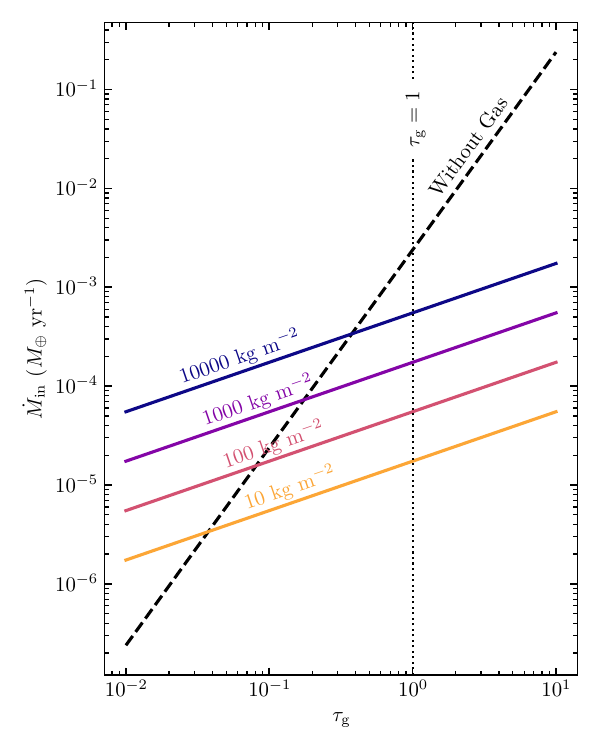}
    \includegraphics[width=0.5\textwidth]{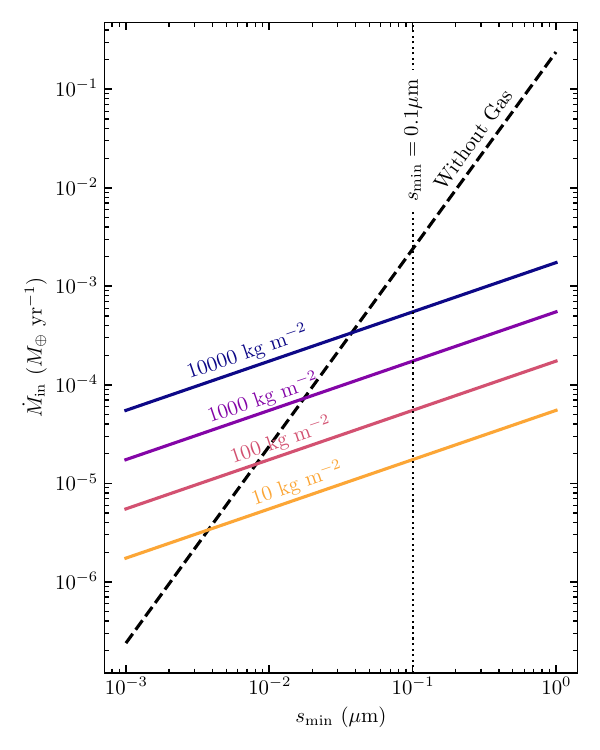}
    \caption{Required mass flux $\dot{M}_\mathrm{in}$ (in earth masses per year) against the cloud's geometric optical depth $\tau_\mathrm{g}$ (left panel). Both the case where there is no gas (solid line), as well as the one with gas (dashed lines) are given. The latter is plotted for different values of the gas surface density, $\Sigma_\mathrm{gas}$. In the right panel the required mass flux, $\dot{M}_\mathrm{in}$ (in earth masses per year), is plotted against the size of the smallest particles, $s_\mathrm{min}$. Both the case where there is no gas (solid line), as well as the one with gas (dashed lines) are given. The latter is plotted for different values of the gas surface density, $\Sigma_\mathrm{gas}$.}
    \label{fig:mass_flux}
\end{figure*}

The equations for the mass flux required for an opaque cloud (derived in Sect. \ref{sec:required_mass_flux}) show the effect of the presence of gas: using our fiducial parameters (see table \ref{tab:variables}), the presence of gas decreases the mass flux by a little over an order of magnitude. More importantly, this mass flux can be brought down further by decreasing the gas density, since $\dot{M}_\mathrm{in}\propto \Sigma_\mathrm{gas}^{0.5}$. This is due to a decreased surface density increasing the time it takes for a planetesimal to be damped into the midplane, causing $s_\mathrm{cc}$ to increase and more particles to be involved in the cascade producing small particles.

In the left panel of Fig. \ref{fig:mass_flux}, we show how the mass flux scales with the optical depth, $\tau_\mathrm{g}$, for multiple gas (surface) densities (solid lines) and when is no gas at all (dashed black line). First of all, we can see that in the absence of gas, the required mass flux scales with $\tau_\mathrm{g}^2$ instead of $\tau_\mathrm{g}^{0.5}$ in the presence of gas. In other words: increasing the mass flux has a stronger effect on the optical depth when there is gas compared to when there is no gas. Increasing the mass flux increases the abundance of particles in the collisional cascade, which decreases their collisional lifetime. In the case where there is no gas present, the lifetime of small particles is their collisional lifetime, and the decline thereof negatively affects the abundance of small particles. The lifetime of small particles in gas, the settling timescale, is unaffected by the mass flux however. Therefore, the effect of increasing the mass flux on the optical depth is stronger when there is gas around; or, equivalently, increasing the optical depth causes a stronger increase in the required mass flux in the gas-free case.

As a result of this for very low optical depth particles become very sparse, increasing their collision times enough to make the no-gas case preferable. At higher optical depths however settling times are much longer than collision times, and having gas around is preferable.  Decreasing the gas surface density by a factor of 1000 has a profound effect and reduces the required mass flux to as little as $10^{-5} M_\oplus$\,yr$^{-1}$, an amount that could be sustainable for millions of years during the disk's lifetime.

The right panel of Fig. \ref{fig:mass_flux} shows how the required mass flux changes when the smallest particle size deviates from $s_\mathrm{min}=0.1$ $\mu$m. Most importantly, having a smaller $s_\mathrm{min}$ decreases the mass flux required for a certain optical depth, since smaller particles have a higher surface area to mass ratio, meaning the same mass is more efficient at absorbing light. It should be noted that we are comparing the gas and no gas cases for equal values of $s_\mathrm{min}$. When there is no gas present, however, small particles are removed almost instantly due to radiation pressure \citep{Backman_PPIII}; whereas these particles would remain in the system when there is gas present. Therefore, the effective value of $s_\mathrm{min}$ would be higher in a system without gas than it is in one with gas, making the latter more preferable.

\subsection{Timescales}
\begin{figure}
    \centering
    \includegraphics[width=\linewidth]{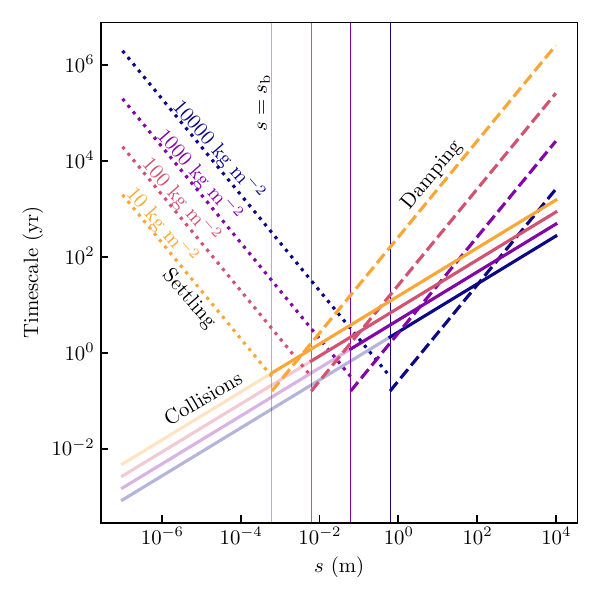}
    \caption{Settling (dotted lines), damping (dashed lines), and collisional (solid lines) timescales for different gas densities. The size below which damping becomes quicker than collisions becomes smaller with decreasing gas density. Small particles below $s_\mathrm{b}$ (thin vertical lines) do not collide since they move with the gas, the settling which occurs instead is much slower.}
    \label{fig:timescales}
\end{figure}
In Fig. \ref{fig:timescales}, we show the timescales at which particles settle or damp into the midplane and at which they collide if they were in the collisional cascade. As described in Sect. \ref{sec:gas_effect}, small particles below $s_\mathrm{b}$ (given by the vertical lines) move with the gas and are thus removed from the cascade. In this figure we can also see how intermediate-size particles ($s_\mathrm{b}<s<s_\mathrm{cc}$), which can collide, are essentially removed from the cascade as well because they are damped into the midplane more quickly than they collide. 

What might not be obvious is why the collisional timescales change with gas density. Since we require an optical depth of 1, a certain amount of small particles is required. For lower gas densities, $s_\mathrm{cc}$ increases, causing relatively smaller particles to be included in the production of small particles. This is more efficient than when only very large particles are producing small particles; therefore, the value of $f_{0,\mathrm{cc}}$ can be lower for the same production rate and the collisional timescales increase.

\subsection{Influence of the fragment size distribution}
\begin{figure}
    \centering
    \includegraphics[width=\linewidth]{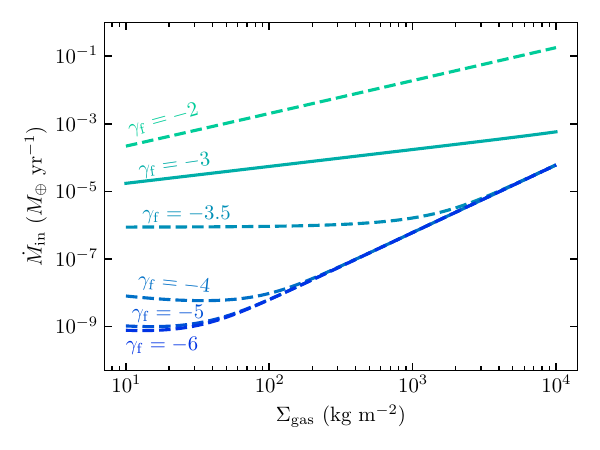}
    \caption{Required mass flux, $\dot{M}_\mathrm{in}$, for an opacity, $\tau_\mathrm{g}$, of 1 as a function of the gas surface density, $\Sigma_\mathrm{gas}$, for different values of the fragment size distribution power, $\gamma_\mathrm{f}$.}
    \label{fig:M_vs_sigma_gamma}
\end{figure}
All the results shown above have the underlying assumption that the fragment size distribution follows a power law with $\gamma_\mathrm{f}=-3$. As described in Sect. \ref{subsec:gamma_f}, this value can be easily modified by working out the integrals numerically instead of algebraically. This value not only influences the required mass flux, but affects how this mass flux changes with gas density as well. This is shown in Fig. \ref{fig:M_vs_sigma_gamma}, where we plot the required mass flux against the gas density for different values of $\gamma_\mathrm{f}$. As expected, lower values of $\gamma_\mathrm{f}$ are more efficient at the production of small particles, reducing the required mass flux significantly.

As we can see, having $\gamma_\mathrm{f}=-6$ instead of $\gamma_\mathrm{f}=-3$ reduces the required mass flux by almost five orders of magnitude for $\Sigma_\mathrm{gas}=1000$ kg m$^{-3}$. This decreases the abundance of particles in the collisional cascade, which increases their collisional timescale as well. Since $f_{0,\mathrm{cc}}\propto \dot{M}_\mathrm{in}^{0.5}$ we find $t_\mathrm{coll}\propto f_{0,\mathrm{cc}}^{-1}\propto \dot{M}_\mathrm{in}^{-0.5}$, which means that decreasing the mass flux by five orders of magnitude increases the collisional timescales by about a factor of $\sqrt{100\textrm{,}000}\approx 300$. Since the collisional timescale of the largest particles in Fig. \ref{fig:timescales} is about 100 yrs, this will become about 30,000 yrs, which is still short enough to expect a steady-state collisional cascade to establish within the disk's lifetime.

\subsection{Caveats}
In deriving our results, some crude assumptions have been made. We assumed the gas to be homogeneous in density, whereas in reality a vertically isothermal disk will have a Gaussian density profile around the midplane \citep{1973A&A....24..337S}. This would cause particles which are just large enough to not bind to the gas high up in the cloud to do so when crossing the midplane. Therefore, these particles were removed from the higher regions of the cloud within a single orbital period. However, in the current picture, these particles would fall in the intermediate-size range, meaning they are damped into the midplane quicker than they collide and did not contribute to the production of smaller particles nor to the cloud's optical depth in the first place. Furthermore, for low values of $\gamma_\mathrm{f}$, the largest particles in the collisional cascade dominate the production of small particles, making these intermediate-size particles even more irrelevant.

We have made another assumption by not taking into account coagulation in the process of settling. As shown by \citet[see][and references therein]{armitage}, this would significantly speed up the process of settling, decreasing the lifetime of small particles, which would increase the required mass flux. However, in the case for a cloud of dust which is just optically thick ($\tau_\mathrm{g}\approx1$), particles would in general not encounter many others on their way into the midplane. In case of higher optical depths, and by extension higher abundance of small particles, coagulation would have to be taken into account.

Furthermore, we have ignored the presence of particles which have
reached the midplane, while these might still act as collision
partners for larger particles, which are still on their inclined orbits
around the central star. This would increase the chance of particles
colliding in the midplane relative to higher up in the cloud, which
would lead to an increased concentration of collisional fragments in
the midplane. Since we require a large part of the star to be covered
by dust grains, this would be detrimental to the efficiency of this
system. However, most particles which end up in the midplane are small
(through settling) or intermediate-size (through damping) and since
only particles larger than $\epsilon s$ are capable of destroying a
particle of a size, $s$, most larger particles which are still on their
inclined orbits will not be affected by an encounter with these
particles.

\section{Conclusions}
\label{sec:conclusions}

We have studied the effect collisions between parent bodies on strongly inclined orbits on the abundance of small grains in planet-forming disks that are in transition and show a decrease in the gas surface density along with significant NIR excess. We anchored our calculations by requiring the optical depth of the collisionally produced dust to be of order 1. Our findings are as follows:
\begin{enumerate}
    \item The presence of gas in the system causes the smallest particles to couple to the gas and stop colliding. Because the time it takes these particles to settle into the midplane is longer than the average time it takes for these particles to collide in a collisional cascade, the presence of gas increases the abundance of small particles significantly compared to a gas-free debris disk.
    \item Since (in the gas-rich case) particles are constantly removed from the higher regions of the cloud by settling, continuous replenishment is needed. Under very conservative assumptions for the fragmentation power law, a mass influx of $2.4\times10^{-3}$ $M_\oplus$/yr is required when there is no gas around and all particles are in the collisional cascade. In the presence of gas with standard values for the gas surface density, the increased lifetime of small particles causes this number to drop by over an order of magnitude to $1.7\times10^{-4}$ $M_\oplus$/yr.
    \item In a transition disk, where the gas surface density is strongly reduced compared to younger disks, the required mass flux is reduced to about $10^{-5}$ $M_\oplus$/yr. This is due to the fact that for higher gas densities, the damping of unbound particles into the midplane is quicker, which increases the particle size below which this process dominates over the process of collisions and fragmentation. Therefore, the production and subsequent abundance of small particles is lower when the gas density is higher.
    \item In deriving these results, we assumed a fragmentation power-law size distribution with $\gamma_\mathrm{f}=-3$. This is a very conservative value, since values down to $-5$ or $-6$ are plausible for the collisional velocities we are dealing with, which would be more efficient at producing small particles. Lower values for this power law lead to a decrease in the required mass flux drastically by several orders of magnitude, down to values as low as $\sim10^{-9}$ $M_\oplus$/yr. Such mass fluxes appear to be easily sustainable for extended periods of time in planet-forming disks.
    \item The timescales at which 10 km-sized particles collide are of the order of 100 yrs for $\gamma_\mathrm{f}=-3$, which could increase to about $30$ kyrs for $\gamma_\mathrm{f}=-6$. These timescales are long, but short enough to expect a steady-state collisional cascade to establish within a disk's lifetime.
\end{enumerate}

In summary, we have shown that a collisional cascade in a
planet-forming disk with reduced but non-zero gas surface densities
can produce observable amounts of dust for extended periods of time. This would make this mechanism a prime candidate for explaining warm IR excess
in transition disks.

\begin{acknowledgements}
This paper has been a long time coming. CD would like to thank Rens Waters, Greet Decin, Rik van Lieshout, Sebastiaan Krijt, Sean Raymond, Myriam Benisty and Kees Dullemond for valuable past discussions and encouragement on the subject of this paper. This research was supported by the International Space Science Institute (ISSI) in Bern, through the ISSI International Team project “Zooming In On Rocky Planet Formation”.
\end{acknowledgements}

\bibliographystyle{aa}
\bibliography{ms}

\begin{thebibliography}{53}
\expandafter\ifx\csname natexlab\endcsname\relax\def\natexlab#1{#1}\fi

\bibitem[{{ALMA Partnership} {et~al.}(2015){ALMA Partnership}, {Brogan},
  {P{\'e}rez}, {Hunter}, {Dent}, {Hales}, {Hills}, {Corder}, {Fomalont},
  {Vlahakis}, {Asaki}, {Barkats}, {Hirota}, {Hodge}, {Impellizzeri}, {Kneissl},
  {Liuzzo}, {Lucas}, {Marcelino}, {Matsushita}, {Nakanishi}, {Phillips},
  {Richards}, {Toledo}, {Aladro}, {Broguiere}, {Cortes}, {Cortes}, {Espada},
  {Galarza}, {Garcia-Appadoo}, {Guzman-Ramirez}, {Humphreys}, {Jung}, {Kameno},
  {Laing}, {Leon}, {Marconi}, {Mignano}, {Nikolic}, {Nyman}, {Radiszcz},
  {Remijan}, {Rod{\'o}n}, {Sawada}, {Takahashi}, {Tilanus}, {Vila Vilaro},
  {Watson}, {Wiklind}, {Akiyama}, {Chapillon}, {de Gregorio-Monsalvo}, {Di
  Francesco}, {Gueth}, {Kawamura}, {Lee}, {Nguyen Luong}, {Mangum}, {Pietu},
  {Sanhueza}, {Saigo}, {Takakuwa}, {Ubach}, {van Kempen}, {Wootten},
  {Castro-Carrizo}, {Francke}, {Gallardo}, {Garcia}, {Gonzalez}, {Hill},
  {Kaminski}, {Kurono}, {Liu}, {Lopez}, {Morales}, {Plarre}, {Schieven},
  {Testi}, {Videla}, {Villard}, {Andreani}, {Hibbard}, \&
  {Tatematsu}}]{ALMA-2015-HLTau}
{ALMA Partnership}, {Brogan}, C.~L., {P{\'e}rez}, L.~M., {et~al.} 2015, \apjl,
  808, L3

\bibitem[{Armitage(2010)}]{armitage}
Armitage, P.~J. 2010, The Astrophysics of Planet Formation (Cambridge
  University Press)

\bibitem[{{Backman} \& {Paresce}(1993)}]{Backman_PPIII}
{Backman}, D.~E. \& {Paresce}, F. 1993, in Protostars and Planets III, ed.
  E.~H. {Levy} \& J.~I. {Lunine}, 1253

\bibitem[{{Benisty} {et~al.}(2011){Benisty}, {Renard}, {Natta}, {Berger},
  {Massi}, {Malbet}, {Garcia}, {Isella}, {M{\'e}rand}, {Monin}, {Testi},
  {Thi{\'e}baut}, {Vannier}, \& {Weigelt}}]{Benisty_2011}
{Benisty}, M., {Renard}, S., {Natta}, A., {et~al.} 2011, \aap, 531, A84

\bibitem[{{Bernab{\`o}} {et~al.}(2022){Bernab{\`o}}, {Turrini}, {Testi},
  {Marzari}, \& {Polychroni}}]{Bernabo_2022}
{Bernab{\`o}}, L.~M., {Turrini}, D., {Testi}, L., {Marzari}, F., \&
  {Polychroni}, D. 2022, \apjl, 927, L22

\bibitem[{{Casassus} {et~al.}(2013){Casassus}, {van der Plas}, {Perez}, {Dent},
  {Fomalont}, {Hagelberg}, {Hales}, {Jord{\'a}n}, {Mawet}, {M{\'e}nard},
  {Wootten}, {Wilner}, {Hughes}, {Schreiber}, {Girard}, {Ercolano}, {Canovas},
  {Rom{\'a}n}, \& {Salinas}}]{Casassus-2013-streamers}
{Casassus}, S., {van der Plas}, G.~M., {Perez}, S., {et~al.} 2013, \nat, 493,
  191

\bibitem[{{D'Alessio} {et~al.}(1999){D'Alessio}, {Calvet}, {Hartmann},
  {Lizano}, \& {Cant{\'o}}}]{dAlessio-accretion-for-excess-1999}
{D'Alessio}, P., {Calvet}, N., {Hartmann}, L., {Lizano}, S., \& {Cant{\'o}}, J.
  1999, \apj, 527, 893

\bibitem[{{D'Alessio} {et~al.}(1998){D'Alessio}, {Cant{\"o}}, {Calvet}, \&
  {Lizano}}]{dAlessio-1998-irradiated-disks}
{D'Alessio}, P., {Cant{\"o}}, J., {Calvet}, N., \& {Lizano}, S. 1998, \apj,
  500, 411

\bibitem[{{D'Angelo} \& {Marzari}(2022)}]{dAngelo-Marzari-2022}
{D'Angelo}, G. \& {Marzari}, F. 2022, \mnras, 509, 3181

\bibitem[{Dohnanyi(1969)}]{dohnanyi}
Dohnanyi, J. 1969, Journal of Geophysical Research, 74, 2531

\bibitem[{Dominik \& Decin(2003)}]{vega_phenomenon}
Dominik, C. \& Decin, G. 2003, the Astrophysical Journal, 598, 626

\bibitem[{{Dominik} {et~al.}(2003{\natexlab{a}}){Dominik}, {Dullemond},
  {Waters}, \& {Natta}}]{DDN-2003}
{Dominik}, C., {Dullemond}, C.~P., {Waters}, L.~B.~F.~M., \& {Natta}, A.
  2003{\natexlab{a}}, in Astronomical Society of the Pacific Conference Series,
  Vol. 287, Galactic Star Formation Across the Stellar Mass Spectrum, ed. J.~M.
  {De Buizer} \& N.~S. {van der Bliek}, 313--318

\bibitem[{{Dominik} {et~al.}(2003{\natexlab{b}}){Dominik}, {Dullemond},
  {Waters}, \& {Walch}}]{Dominik-2003-passive-disk-model}
{Dominik}, C., {Dullemond}, C.~P., {Waters}, L.~B.~F.~M., \& {Walch}, S.
  2003{\natexlab{b}}, \aap, 398, 607

\bibitem[{{Dr{\k{a}}{\.z}kowska} {et~al.}(2023){Dr{\k{a}}{\.z}kowska},
  {Bitsch}, {Lambrechts}, {Mulders}, {Harsono}, {Vazan}, {Liu}, {Ormel},
  {Kretke}, \& {Morbidelli}}]{Drazkowska-PPVII}
{Dr{\k{a}}{\.z}kowska}, J., {Bitsch}, B., {Lambrechts}, M., {et~al.} 2023, in
  Astronomical Society of the Pacific Conference Series, Vol. 534, Protostars
  and Planets VII, ed. S.~{Inutsuka}, Y.~{Aikawa}, T.~{Muto}, K.~{Tomida}, \&
  M.~{Tamura}, 717

\bibitem[{{Dullemond} \& {Dominik}(2004)}]{DD-2004-flaring-shadow}
{Dullemond}, C.~P. \& {Dominik}, C. 2004, \aap, 417, 159

\bibitem[{{Hillenbrand} {et~al.}(1992){Hillenbrand}, {Strom}, {Vrba}, \&
  {Keene}}]{Hillenbrand-1992}
{Hillenbrand}, L.~A., {Strom}, S.~E., {Vrba}, F.~J., \& {Keene}, J. 1992, \apj,
  397, 613

\bibitem[{{Kennedy} \& {Wyatt}(2010)}]{2010MNRAS.405.1253K}
{Kennedy}, G.~M. \& {Wyatt}, M.~C. 2010, \mnras, 405, 1253

\bibitem[{{Kenyon} \& {Bromley}(2005)}]{2005AJ....130..269K}
{Kenyon}, S.~J. \& {Bromley}, B.~C. 2005, \aj, 130, 269

\bibitem[{{Kenyon} \& {Bromley}(2008)}]{2008ApJS..179..451K}
{Kenyon}, S.~J. \& {Bromley}, B.~C. 2008, \apjs, 179, 451

\bibitem[{{Kenyon} \& {Bromley}(2010)}]{2010ApJS..188..242K}
{Kenyon}, S.~J. \& {Bromley}, B.~C. 2010, \apjs, 188, 242

\bibitem[{{Kobayashi} \& {L{\"o}hne}(2014)}]{2014MNRAS.442.3266K}
{Kobayashi}, H. \& {L{\"o}hne}, T. 2014, \mnras, 442, 3266

\bibitem[{{Kobayashi} \& {Tanaka}(2010)}]{2010Icar..206..735K}
{Kobayashi}, H. \& {Tanaka}, H. 2010, \icarus, 206, 735

\bibitem[{{Krijt} \& {Dominik}(2011)}]{2011A&A...531A..80K}
{Krijt}, S. \& {Dominik}, C. 2011, \aap, 531, A80

\bibitem[{{Krivov} {et~al.}(2006{\natexlab{a}}){Krivov}, {L{\"o}hne}, \&
  {Srem{\v{c}}evi{\'c}}}]{2006A&A...455..509K}
{Krivov}, A.~V., {L{\"o}hne}, T., \& {Srem{\v{c}}evi{\'c}}, M.
  2006{\natexlab{a}}, \aap, 455, 509

\bibitem[{{Krivov} {et~al.}(2006{\natexlab{b}}){Krivov}, {L{\"o}hne}, \&
  {Srem{\v{c}}evi{\'c}}}]{Krivov_model_2006}
{Krivov}, A.~V., {L{\"o}hne}, T., \& {Srem{\v{c}}evi{\'c}}, M.
  2006{\natexlab{b}}, \aap, 455, 509

\bibitem[{{Krivov} {et~al.}(2005){Krivov}, {Srem{\v{c}}evi{\'c}}, \&
  {Spahn}}]{Krivov_basics_2005}
{Krivov}, A.~V., {Srem{\v{c}}evi{\'c}}, M., \& {Spahn}, F. 2005, \icarus, 174,
  105

\bibitem[{{Krivov} \& {Wyatt}(2021)}]{Krivov_Wyatt_2020}
{Krivov}, A.~V. \& {Wyatt}, M.~C. 2021, \mnras, 500, 718

\bibitem[{{Lagrange} {et~al.}(2000){Lagrange}, {Backman}, \&
  {Artymowicz}}]{Lagrange-PPVI}
{Lagrange}, A.~M., {Backman}, D.~E., \& {Artymowicz}, P. 2000, in Protostars
  and Planets IV, ed. V.~{Mannings}, A.~P. {Boss}, \& S.~S. {Russell}, 639

\bibitem[{{Lissauer} \& {Stewart}(1993)}]{Lissauer_Steward_PPIII}
{Lissauer}, J.~J. \& {Stewart}, G.~R. 1993, in Protostars and Planets III, ed.
  E.~H. {Levy} \& J.~I. {Lunine}, 1061

\bibitem[{{L{\"o}hne} {et~al.}(2008){L{\"o}hne}, {Krivov}, \&
  {Rodmann}}]{2008ApJ...673.1123L}
{L{\"o}hne}, T., {Krivov}, A.~V., \& {Rodmann}, J. 2008, \apj, 673, 1123

\bibitem[{{Maaskant} {et~al.}(2013){Maaskant}, {Honda}, {Waters}, {Tielens},
  {Dominik}, {Min}, {Verhoeff}, {Meeus}, \& {van den
  Ancker}}]{Maaskant-2013-group-I-transitional}
{Maaskant}, K.~M., {Honda}, M., {Waters}, L.~B.~F.~M., {et~al.} 2013, \aap,
  555, A64

\bibitem[{{Meeus} {et~al.}(2001){Meeus}, {Waters}, {Bouwman}, {van den Ancker},
  {Waelkens}, \& {Malfait}}]{Meeus-2001-groups}
{Meeus}, G., {Waters}, L.~B.~F.~M., {Bouwman}, J., {et~al.} 2001, \aap, 365,
  476

\bibitem[{{Natta} {et~al.}(2001{\natexlab{a}}){Natta}, {Prusti}, {Neri},
  {Wooden}, {Grinin}, \& {Mannings}}]{Natta-2001-inner-disk-edge}
{Natta}, A., {Prusti}, T., {Neri}, R., {et~al.} 2001{\natexlab{a}}, \aap, 371,
  186

\bibitem[{{Natta} {et~al.}(2001{\natexlab{b}}){Natta}, {Prusti}, {Neri},
  {Wooden}, {Grinin}, \& {Mannings}}]{Natta_2001}
{Natta}, A., {Prusti}, T., {Neri}, R., {et~al.} 2001{\natexlab{b}}, \aap, 371,
  186

\bibitem[{O'Brien \& David(2003)}]{OBrien}
O'Brien \& David, P. 2003, Icarus, 164, 334

\bibitem[{{Pinilla} {et~al.}(2016){Pinilla}, {Klarmann}, {Birnstiel},
  {Benisty}, {Dominik}, \& {Dullemond}}]{Pinilla-2016-tunnel}
{Pinilla}, P., {Klarmann}, L., {Birnstiel}, T., {et~al.} 2016, \aap, 585, A35

\bibitem[{{Price} {et~al.}(2018){Price}, {Cuello}, {Pinte}, {Mentiplay},
  {Casassus}, {Christiaens}, {Kennedy}, {Cuadra}, {Sebastian Perez}, {Marino},
  {Armitage}, {Zurlo}, {Juhasz}, {Ragusa}, {Laibe}, \&
  {Lodato}}]{price-2018-streamers}
{Price}, D.~J., {Cuello}, N., {Pinte}, C., {et~al.} 2018, \mnras, 477, 1270

\bibitem[{{Raymond} {et~al.}(2011){Raymond}, {Armitage}, {Moro-Mart{\'\i}n},
  {Booth}, {Wyatt}, {Armstrong}, {Mandell}, {Selsis}, \&
  {West}}]{2011A&A...530A..62R}
{Raymond}, S.~N., {Armitage}, P.~J., {Moro-Mart{\'\i}n}, A., {et~al.} 2011,
  \aap, 530, A62

\bibitem[{{Raymond} {et~al.}(2012){Raymond}, {Armitage}, {Moro-Mart{\'\i}n},
  {Booth}, {Wyatt}, {Armstrong}, {Mandell}, {Selsis}, \&
  {West}}]{2012A&A...541A..11R}
{Raymond}, S.~N., {Armitage}, P.~J., {Moro-Mart{\'\i}n}, A., {et~al.} 2012,
  \aap, 541, A11

\bibitem[{{Raymond} \& {Izidoro}(2017)}]{Raymond_Izidoro_2017}
{Raymond}, S.~N. \& {Izidoro}, A. 2017, \icarus, 297, 134

\bibitem[{{Shakura} \& {Sunyaev}(1973)}]{1973A&A....24..337S}
{Shakura}, N.~I. \& {Sunyaev}, R.~A. 1973, \aap, 24, 337

\bibitem[{Takasawa {et~al.}(2011)Takasawa, Nakamura, Kadono, Arakawa, Dohi,
  Ohno, Seto, Maeda, Shigemori, Hironaka, Sakaiya, Fujioka, Sano, Otani,
  Watari, Sangen, Setoh, Machii, \& Takeuchi}]{Takasawa_2011}
Takasawa, S., Nakamura, A.~M., Kadono, T., {et~al.} 2011, The Astrophysical
  Journal Letters, 733, L39

\bibitem[{Tanaka {et~al.}(1996)Tanaka, Inaba, \&
  Nakazawa}]{tanaka_1996_steadystate}
Tanaka, H., Inaba, S., \& Nakazawa, K. 1996, Icarus, 123, 450

\bibitem[{{Th{\'e}bault} \& {Augereau}(2007)}]{2007A&A...472..169T}
{Th{\'e}bault}, P. \& {Augereau}, J.~C. 2007, \aap, 472, 169

\bibitem[{{Th{\'e}bault} {et~al.}(2003){Th{\'e}bault}, {Augereau}, \&
  {Beust}}]{2003A&A...408..775T}
{Th{\'e}bault}, P., {Augereau}, J.~C., \& {Beust}, H. 2003, \aap, 408, 775

\bibitem[{{Turrini} {et~al.}(2019){Turrini}, {Marzari}, {Polychroni}, \&
  {Testi}}]{Turrini-2019T}
{Turrini}, D., {Marzari}, F., {Polychroni}, D., \& {Testi}, L. 2019, \apj, 877,
  50

\bibitem[{{Valeg{\r{a}}rd} {et~al.}(2021){Valeg{\r{a}}rd}, {Waters}, \&
  {Dominik}}]{Valegard-2021-IMTT}
{Valeg{\r{a}}rd}, P.~G., {Waters}, L.~B.~F.~M., \& {Dominik}, C. 2021, \aap,
  652, A133

\bibitem[{{van Lieshout} {et~al.}(2014){van Lieshout}, {Dominik}, {Kama}, \&
  {Min}}]{2014A&A...571A..51V}
{van Lieshout}, R., {Dominik}, C., {Kama}, M., \& {Min}, M. 2014, \aap, 571,
  A51

\bibitem[{{Wyatt}(2008)}]{2008ARA&A..46..339W}
{Wyatt}, M.~C. 2008, \araa, 46, 339

\bibitem[{{Wyatt} {et~al.}(2011){Wyatt}, {Clarke}, \&
  {Booth}}]{2011CeMDA.111....1W}
{Wyatt}, M.~C., {Clarke}, C.~J., \& {Booth}, M. 2011, Celestial Mechanics and
  Dynamical Astronomy, 111, 1

\bibitem[{{Wyatt} \& {Dent}(2002)}]{Wyatt_Dent_2002}
{Wyatt}, M.~C. \& {Dent}, W.~R.~F. 2002, \mnras, 334, 589

\bibitem[{{Wyatt} {et~al.}(2007{\natexlab{a}}){Wyatt}, {Smith}, {Greaves},
  {Beichman}, {Bryden}, \& {Lisse}}]{2007ApJ...658..569W}
{Wyatt}, M.~C., {Smith}, R., {Greaves}, J.~S., {et~al.} 2007{\natexlab{a}},
  \apj, 658, 569

\bibitem[{{Wyatt} {et~al.}(2007{\natexlab{b}}){Wyatt}, {Smith}, {Su}, {Rieke},
  {Greaves}, {Beichman}, \& {Bryden}}]{2007ApJ...663..365W}
{Wyatt}, M.~C., {Smith}, R., {Su}, K.~Y.~L., {et~al.} 2007{\natexlab{b}}, \apj,
  663, 365

\end{thebibliography}

\begin{appendix} 

\section{Time dependence of the size distribution}
\label{ap:ODE}
To solve Eq. \eqref{eq:size_dist_ODE} for the size distribution of tightly coupled particles, we can substitute:
\begin{equation}
    u\equiv \phi_\textrm{prod}-\frac{f_\textrm{b}}{t_\textrm{settle}},
\end{equation}
with
\begin{equation}
    \frac{\textrm{d}u}{\textrm{d}t}=-\frac{1}{t_\textrm{settle}}\frac{\textrm{d}f_\textrm{b}}{\textrm{d}t},
\end{equation}
to rewrite the differential equation as:
\begin{equation}
    \frac{\textrm{d}u}{u}=-\frac{\textrm{d}t}{t_\textrm{settle}}.
\end{equation}
This can be solved by integration, resulting in:
\begin{equation}
    u(t)=u(t_0)e^{-(t-t_0)/t_\textrm{settle}}.
\end{equation}
Substituting the definition of $u$ again we find
\begin{equation}
    \phi_\textrm{prod}-\frac{f_\textrm{b}(t)}{t_\textrm{settle}}=\left(\phi_\textrm{prod}-\frac{f_\textrm{b}(t_0)}{t_\textrm{settle}}\right)e^{-(t-t_0)/t_\textrm{settle}}.
\end{equation}
And solving for $f_\mathrm{b}$ finds
\begin{equation}
    f_\textrm{b}(t)=t_\textrm{settle}\phi_\textrm{prod}-\left(t_\textrm{settle}\phi_\textrm{prod}-f_\textrm{b}(t_0)\right)e^{-(t-t_0)/t_\textrm{settle}}.
\end{equation}
Now, in the limit where $t\gg t_\textrm{settle}$, the steady-state size distribution is:
\begin{equation}
    f_\textrm{b}(s)=t_\textrm{settle}(s)\phi_\textrm{prod}(s).
\end{equation}

\section{Dust settling and inclination damping}
\label{ap:settling}
A particle above the midplane, suspended in a gas, experiences drag. Depending on the situation, the particle moves in either of two possible ways: (1) the particle is large enough to move through the gas and follow its inclined orbit around the central star. Over a multitude of orbits, drag causes the inclination to decline; (2) the particle couples to the gas, which itself moves parallel to the midplane, and starts to move with it. Gravity slowly pulls the particle downwards, causing it to settle into the midplane.

Only considering the forces in the (vertical) $z$ direction, the acceleration resulting from drag is given by dividing the Epstein drag force (Eq. \ref{eq:epstein}) over the particle mass (Eq. \ref{eq:mass}), resulting in:
\begin{align}
    \frac{F_\textrm{D}}{m}&=-\frac{\frac{4\pi}{3}\rho_\textrm{gas}s^2v_\textrm{th}v}{\frac{4\pi}{3}\rho_\bullet s^3},\\
    &=-\frac{\rho_\textrm{gas}}{\rho_\bullet}\frac{v_\textrm{th}}{s}v.
\end{align}
Furthermore, as described by \cite{armitage}, in the limit where $z\ll R$, the gravitational acceleration in $z$ direction is given by:
\begin{align}
    \frac{F_\textrm{g}}{m}&=-\frac{\mu}{R^2}\frac{z}{R}\\
    &=-\Omega_\textrm{K}^2z.
\end{align}
Using this, the particle's height above the midplane, $z,$ is given by:
\begin{equation}
    \ddot{z}+\frac{\rho_\textrm{gas}}{\rho_\bullet}\frac{v_\textrm{th}}{s}\dot{z}+\Omega_\textrm{K}^2z=0.
\end{equation}
Substituting $t_\textrm{fric}$ from Eq. \ref{eq:t_fric}, this evaluates to
\begin{equation}
    \ddot{z}+t_\textrm{fric}^{-1}\dot{z}+\Omega_\textrm{K}^2z=0.
\end{equation}
Now using $z=A\exp{rt}$ results in a quadratic equation for $r$, which has the solutions
\begin{align}
\label{eq:r_sol}
    r&=\frac{-t_\textrm{fric}^{-1}\pm\sqrt{t_\textrm{fric}^{-2}-4\Omega_\textrm{K}^2}}{2},\\
    &=-(2t_\textrm{fric})^{-1}\pm\sqrt{(2t_\textrm{fric})^{-2}-\Omega_\textrm{K}^2}.
\end{align}
We can now distinguish the two different cases by the sign of the square root's argument. The argument is negative if $t_\textrm{fric}\Omega_\textrm{K}>1/2$, which corresponds closely to the requirement for a particle not being bound to the gas given by \cite{armitage} (which uses $t_\mathrm{fric}\Omega_\mathrm{K}>1$).  In this case an oscillating motion occurs in the $z$-direction: the particle follows it's inclined orbit and periodically crosses the midplane. The angular frequency of this motion is $\sqrt{\Omega_\mathrm{K}^2-(2t_\mathrm{fric})^{-2}}$, meaning the particle's orbit is slightly slowed down relative to the Keplerian orbital frequency $\Omega_\mathrm{K}$. The amplitude of this oscillation, corresponding to the orbit's inclination, is exponentially decreases on a timescale:
\begin{equation}
    t_\textrm{damp}=2t_\textrm{fric}.
\end{equation}
In other words, a particle which is not tightly coupled to the gas is damped into the midplane on a timescale twice the friction timescale.

The other case, where $t_\textrm{fric}\Omega_\textrm{K}<1/2$, describes the settling of a particle into the midplane: the particle has left it's inclined orbit, and instead moves with the gas while slowly approaching the midplane. The timescales at which this happens, given by $-1/r$ from Eq. \eqref{eq:r_sol}, are:
\begin{equation}
    -\frac{1}{r}=\frac{2t_\textrm{fric}}{1\pm\sqrt{1-(2t_\textrm{fric}\Omega_\textrm{K})^2}}.
\end{equation}
In the case where $t_\textrm{fric}\Omega_\textrm{K}\ll 1/2$ (corresponding to a tight coupling to the gas), this can be expressed as:
\begin{align}
    -\frac{1}{r}&=\frac{2t_\textrm{fric}}{1\pm\left(1-2t_\textrm{fric}^2\Omega_\textrm{K}^2\right)}\\
    &=\frac{t_\textrm{fric}}{1-t_\textrm{fric}^2\Omega_\textrm{K}^2} \quad \lor \quad \frac{1}{t_\textrm{fric}\Omega_\textrm{K}^2}.
\end{align}
Since the complete solution is a linear combination of these two exponential functions, the slowest exponential decay will dominate on longer timescales. For $t_\mathrm{fric}\Omega_\mathrm{K}<1/\sqrt{2}$, the second is the largest, so settling occurs on a timescale
\begin{equation}
    t_\textrm{settle}=\frac{1}{t_\textrm{fric}\Omega_\textrm{K}^2}.
\end{equation}
Using $t_\textrm{fric}$ from Eq. \eqref{eq:t_fric}, this evaluates to the settling time as given by Eq. \eqref{eq:t_settle}.

\onecolumn
\section{List of symbols}
This appendix contains the list of symbols used in this paper.
\begin{table}[h]
\caption{All symbols used in this paper.}
\begin{tabular}{ c l l }
    \label{tab:variables}
    \textbf{Symbol} & \textbf{Description} & \textbf{Value$^{(1)}$}\\
    
    \hline
    Dust particles \\
    \hline
    $s$ & particle size &  \\
    $s_\mathrm{min}$ & size of the smallest particles & 0.1 $\mathrm{\mu}$m\\
    $s_\mathrm{b}$ & size below which particles become bound to the gas & Eq. \eqref{eq:s_b}\\
    $s_\mathrm{cc}$ & size below which the collisional lifetime is longer than the damping timescale & Eq. \eqref{eq:s_cc} \\
    $s_\mathrm{pt}$ & size of planetesimals provided to the system & \\
    $m$ & mass of a particle &  Eq. \eqref{eq:mass}\\
    $\rho_\bullet$ & material density of dust and larger particles & 1500 kg m$^{-2}$\\
    $\Omega_\mathrm{K}$ & Kepler frequency & $\sqrt{GM_*/R^3}$\\
    $v_\mathrm{K}$ & Kepler velocity & $\sqrt{GM_*/R}$\\
    $\nu$ & the relative velocity between particle relative to $v_\mathrm{K}$ & $\eta/4$\\
    $\sigma_\mathrm{tot}$ & total collisional cross-section of a particle & Eq. \eqref{eq:sigma_tot}\\
    $\sigma_\mathrm{g}$ & geometric cross-section of a particle & Eq. \eqref{eq:sigma_g}\\
    $\epsilon$ & relative size of the smallest particle capable of destroying particle of a given size & 0.043\\
    $\epsilon_0$ &  & $-\pi\epsilon^{\gamma+1}/(\gamma+1)$\\
    $S$ & material strength/binding energy & 200 J kg$^{-1}$\\
    $t_\mathrm{coll}$ & collisional lifetime of a particle &  Eq. \eqref{eq:t_coll}\\
    $z$ & height of a particle above the midplane &  \\
    
    \hline
    Size distributions \\
    \hline
    $f$ & size distribution & $f(s)=f_0s^\gamma$\\
    $f_0$ & size distribution scaling constant & \\
    $\gamma$ & power-law value for size distributions &  \\
    $f_\mathrm{cc}$, $f_{0,\mathrm{cc}}$ & size distribution of particles in the collisional cascade& Eq. \eqref{eq:f_0cc}\\
    $\gamma_\mathrm{cc}$ & power-law value for steady-state collisional cascade & $-3.5$\\
    $f_\mathrm{int}$, $f_{0,\mathrm{int}}$ & size distribution of intermediate-size particles & Eq. \eqref{eq:f_0cc}\\
    $f_\mathrm{b}$, $f_{0,\mathrm{b}}$ & size distribution of bound particles & Eq. \eqref{eq:f_b_steady_state}\\
    $f_\mathrm{frag}$, $f_{0,\mathrm{frag}}$ & size distribution of fragments resulting from the destruction of a larger particle & Eq. \eqref{eq:fragmentation_integral}\\
    $\gamma_\mathrm{f}$ & power-law value for fragment size distribution & $-3$\\
    $\phi_\mathrm{prod}$ & production rate of particles & Eq. \eqref{eq:phi_prod_integral}\\

    \hline
    Dust cloud\\
    \hline
    $R$ & distance of the dust cloud to the star & 1 au\\
    $\Delta R$ & width of the dust cloud measured along a line-of-sight from the star & 0.1 au\\
    $\eta$ & fraction of the star surrounded by dust & 0.3\\
    $\tau_\mathrm{g}$ & geometric optical depth of the cloud of dust & 1\\
    $V$ & dust cloud volume & Eq. \eqref{eq:volume}\\
    
    \hline
    Gas \& friction\\
    \hline
    $\Sigma_\mathrm{gas}$ & gas surface density & 1000 kg m$^{-2}$\\
    $T_\mathrm{gas}$ & gas temperature & 300 K\\
    $\mu_\mathrm{mol}$ & mean molecular mass & 2.3\\
    $\rho_\mathrm{gas}$ & gas density & Eq. \eqref{eq:rho_gas}\\
    $v_\mathrm{th}$ & thermal velocity of gas molecules & Eq. \eqref{eq:v_th}\\
    $F_\mathrm{D}$ & drag force & Eq. \eqref{eq:epstein}\\
    $t_\mathrm{fric}$ & friction time of a particle in gas & Eq. \eqref{eq:t_fric}\\
    $\tau_\mathrm{fric}$ & relative friction time of a particle in gas & Eq. \eqref{eq:tau_fric}\\
    $t_\mathrm{settle}$ & settling time of a particle & Eq. \eqref{eq:t_settle}\\
    
    \hline
    General\\
    \hline
    $\dot{M}_\mathrm{in}$ & mass flux of planetesimals into the system & Eq. \eqref{eq:M_in_from_f_cc}\\
    $M_*$ & mass of central star & 1 $M_\odot$\\
    $G$ & gravitational constant & $6.67\times 10^{-11}$ N m$^2$ kg$^{-2}$\\
    $k_\mathrm{B}$ & Boltzmann constant & $1.38\times10^{-23}$ J K$^{-1}$\\
    $m_\mathrm{p}$ & proton mass & $1.67\times10^{-27}$ kg\\
\end{tabular}
$^{(1)}$ For free parameters, the fiducial value is given. For secondary variables, the equation is given.
\end{table}
\end{appendix}

\end{document}